\documentclass[graybox]{svmult}

\usepackage[round]{natbib}
\usepackage{mathptmx}       
\usepackage{helvet}         
\usepackage{courier}        
\usepackage{type1cm}        
\usepackage{makeidx}         
\usepackage{graphicx}        
\usepackage{multicol}        
\usepackage[bottom]{footmisc}

\makeindex             

\def\Rsun{\ifmmode{R_\odot}\else{R$_\odot$}\fi}

\begin{document}

\title*{Coronal cavities: observations and implications for the magnetic environment of prominences}
\titlerunning{Coronal cavities} 
\author{Sarah Gibson}

\institute{Sarah Gibson \at HAO/NCAR, p.o. box 3000, Boulder, CO, 80301, \email{sgibson@ucar.edu}}

%
%

\maketitle

\abstract{Dark and elliptical, coronal cavities yield important clues to the magnetic structures that cradle prominences, and to the forces that ultimately lead to their eruption. We review observational analyses of cavity morphology, thermal properties (density and temperature), line-of-sight and plane-of-sky flows, substructure including hot cores and central voids, linear polarization signatures, and observational precursors and predictors of eruption.  We discuss a magnetohydrodynamic interpretation of these observations which argues that the cavity is a magnetic flux rope, and pose a set of open questions for further study.}

\section{Introduction}
\label{sec:intro}

The twisted or sheared magnetic field associated with prominences represents stored magnetic energy that may be explosively released in coronal mass ejections (CMEs).  It is therefore essential to establish the full three-dimensional (3D) nature of this magnetic field, in order to be alert to topologies prone to eruption and/or approaching thresholds of instability \citep{fan14_book}.  However, the prominence itself traces only a portion of the 3D field, being generally localized to a narrow structure above the neutral line \citep{engvold14_book}.  For this reason, coronal prominence cavities, which represent a much larger volume than that of the prominence, provide a compelling avenue for exploring the larger-scale magnetic structure associated with the prominence.

\begin{figure}[t!]
\center{\includegraphics[scale=.47]{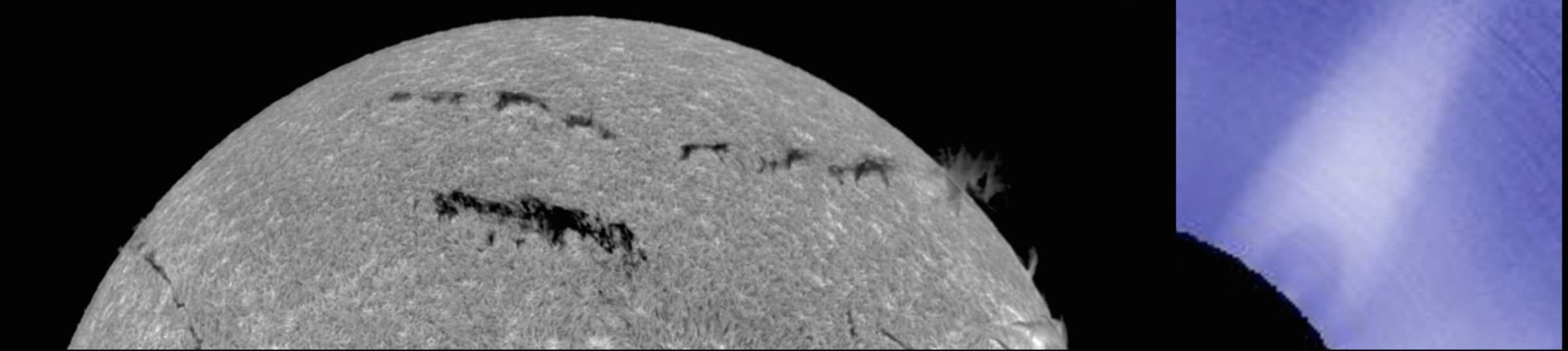}
\includegraphics[scale=.45]{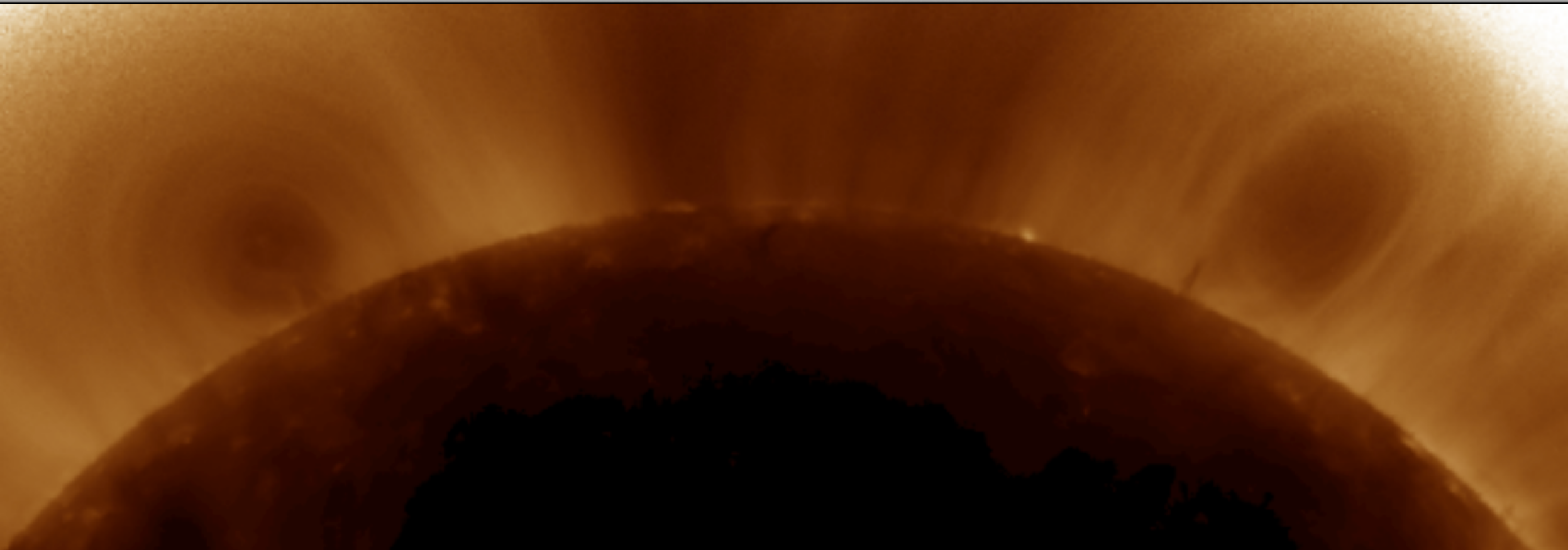}}
\caption{Polar crown filament (PCF) cavities present optimal viewing conditions.  Top: July 22 2002 PCF and associated cavity, (left) as seen in H$\alpha$ by Big Bear Solar Observatory (BBSO) and (right) in white light by Mauna Loa Solar Observatory Mk4 (MLSO/Mk4).  Bottom: Example of PCF cavity manifesting on both limbs, as seen by Solar Dynamics Observatory/Atmospheric Imaging Assembly (SDO/AIA) in $193~\AA$, processed with a radial gradient filter (described in \citep{forland_13}).} 
\label{fig:promcav}       
\end{figure}

\section{Background}
\label{sec:back}

Coronal cavities extend, tunnel-like, above photospheric neutral lines, and are usually observed as dark ellipses or partial ellipses at the limb (Figure \ref{fig:promcav}).  They are associated with prominences in a manner similar to how filament channels are associated with filaments \citep{engvold14_book}: they are often seen surrounding the prominence at the limb, but clear cavities may exist even when no prominence material can be seen.  Cavities may last for days or even weeks \citep{gibcav}, but even when not erupting, like prominences \citep{kucera14_book}, they have flows associated with them \citep{BergerT.promin-plume.2008ApJ...676L..89B,schmit_09,li_12}.   Cavities are most visible when they are lined up along the line of sight, with few neighboring bright features such as active regions to obscure them.  Therefore, polar crown filament (PCF) cavities are particularly good candidates for study.

Cavities can be seen during solar eclipses, so they were initially observed in white light \citep{waldmeier_70,saitohyder,saitotand_73}. Radio observations provided further evidence that cavities were density depletions \citep{straka,kundu}. As space observations became possible, they were also observed in soft X-ray (SXR) \citep{vaiana_73,mcintosh_76,serio_78} and extreme ultraviolet (EUV) \citep{schmahl_79}.  Helium 10830 observations further demonstrated their links to prominences \citep{mccabemickey_81}.  See \citet{tand74,eng89,tand95,gibcav} for reviews of these early observations.

In this paper we will present current understanding of coronal cavities, using multiwavelength observations and MHD theoretical interpretations of prominence/cavity systems.  We will apply these data and models to address two questions in particular: What is the nature of the pre-eruption MHD equilibria represented by quiescent cavities, and what drives these structures to erupt?

\section{Quiescent cavities: MHD equilibria of energized fields}
\label{sec:quiescent}

We will begin by summarizing observations of cavities not in eruption.  These show that quiescent cavities are ubiquitous, possess prolate-elliptical-arched-cylindrical (croissant-like) morphology, have low density (about a factor of two depleted relative to surrounding streamers), are multithermal, demonstrate flows spatially and temporally linking cavity to prominence, have substructure tracing nested ellipses, and possess a ``lagomorphic'' (rabbit-head-shaped) polarimetric signature indicating twisted or sheared magnetic field well above the height of the filament.

As we present these observations, we will discuss how they might be interpreted in the context of MHD theory.  A magnetic flux rope has been proposed as a model for the cavity \citep{lowhund}, and indeed, we will demonstrate that this interpretation is basically consistent with all of the observations.  Nevertheless, ambiguities and open questions remain and we will comment on these throughout this review.   

\subsection{Location and prevalence}
\label{subsec:ubiqobs}

Coronal cavities are not hard to find, except perhaps at solar maximum.  Larger cavities may extend as high as $1.6 \Rsun$ ($0.6 \Rsun$ above the solar surface) \citep{gibcav,fuller_09}, and these are best observed in white light.  Smaller cavities are better observed in the emission corona (e.g., EUV, SXR) as they may partly or wholly lie below the occulting disk of white-light coronagraphs.  

The frequency of observed cavities was much greater in a survey of EUV cavities \citep{forland_13} than in a survey of white light cavities \citep{gibcav}: nearly $80\%$ of days surveyed had one or more EUV cavity, vs. $~10\%$ of white light days surveyed.  This was due in part to the greater data coverage of a space-based coronal imager (no occulting disk, 24-hour coverage) compared to a ground-based coronagraph, but also to the phase of the solar cycle.  The white-light survey was centered around solar maximum, when the global corona is most complex and cavities are likely to be obscured by surrounding bright features.  On the other hand, the EUV study was undertaken during the ascending phase of the cycle, which is an excellent time for viewing cavities.  The presence of PCFs in both hemispheres meant that cavities were frequently found in multiple solar quadrants (e.g., Figure \ref{fig:promcav}, bottom panel).

\runinhead{MHD interpretation of cavity ubiquity: minimum energy states that conserve helicity.}
As discussed in \citet{fan14_book}, prominences are generally modeled in terms of twisted or sheared magnetic fields.  The non-potentiality of such fields represents stored magnetic energy, and accumulated magnetic helicity.  Magnetic energy can be reduced on relatively short time scales through the spontaneous formation and dissipation of current sheets \citep{parker,janse_10}.  This mode of dissipation is subject, under condition of high electrical conductivity, to the magnetic helicity remaining approximately conserved as a global quantity \citep{taylor_74,Berger84}.  It is therefore likely that, over time, excess magnetic energy that can be dissipated will be, resulting in a lower energy state that maintains an accumulated helicity. 

For a given boundary condition, the minimum energy state conserving helicity is that of a constant-$\alpha$ force-free field \citep{woltjer_58}.   For sufficiently large values of helicity, the constant-$\alpha$ force-free solution is a magnetic flux rope: a coherent structure in which magnetic field lines wind about a central axis \citep{low_94}. PCF evolution is likely to involve a steady accumulation of helicity over weeks and even months, resulting in the formation of stable magnetic flux ropes \citep{mackay14_book}.  Thus, a characteristic flux rope topology that creates a cavity in combination with a geometry conducive to unobstructed viewing may explain cavity clarity and prevalence in PCFs.

\begin{svgraybox}
\runinhead{Open Questions}
\begin{itemize}
\item Do all prominences have cavities extending above them, with only a subset unobscured?
\item Does the presence (or lack) of a cavity depend on whether sufficient helicity can accumulate to form a stable flux rope?  
\end{itemize}
\end{svgraybox}

\begin{figure}[b!]
\center{\includegraphics[scale=.4]{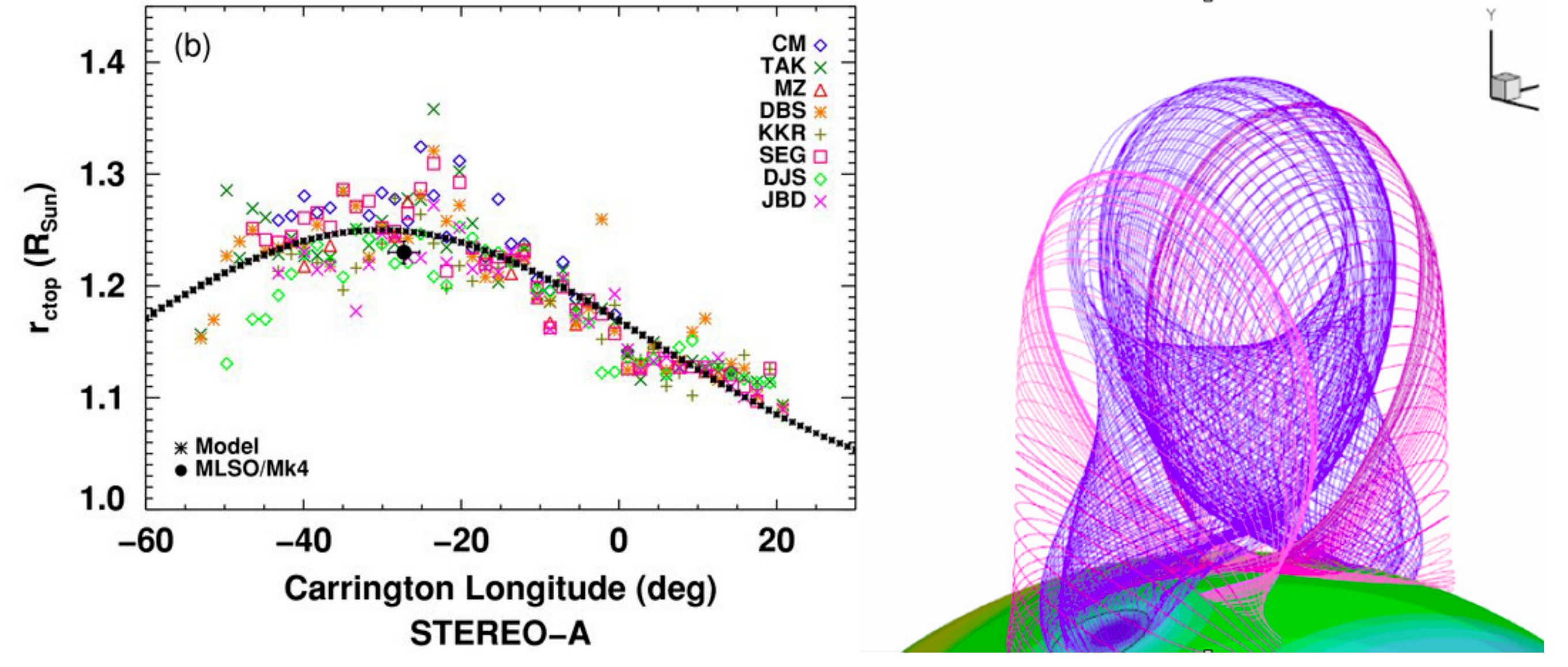}}
\caption{Cavities have arched, tunnel-like morphologies with elliptical cross-sections. Left: Multiple-observer measurements of cavity ellipse height vs. longitude/date (from \citet{gibson_10}, reproduced by permission of the AAS). Right: flux surfaces of a magnetic flux rope simulation (described in \citep{gibfan_06b}).}
\label{fig:croissant}       
\end{figure}

\subsection{3D morphology}
\label{subsec:morphobs}

In analyzing cavities it is critical to keep in mind that they are optically thin, unlike the prominences embedded in them.  
 It is straightforward to measure the 3D morphology of EUV cavities, which can be observed down to the limb of the Sun.  In contrast,  white light cavities usually have their bottom portions obscured by the occulting disk of a coronagraph. 
The plane-of-sky projection of EUV cavities are generally well fit by ellipses. The aspect ratio for cavity ellipses shows a strong tendency towards prolateness: $93\%$ of 119 EUV cavities surveyed were taller than they were wide \citep{forland_13}.  

Measuring their three-dimensional extent requires multiple viewing angles and/or an observation of the structure rotating past the plane of the sky (under the assumption that the structure does not change).  The extension of the cavity above the neutral line is essentially a cylindrical structure with a Gaussian variation in height (Figure \ref{fig:croissant} left) \citep{gibson_10}. Therefore the 3D morphology of cavities may be characterized as that of a prolate-elliptical-arched cylinder, or more familiarly as the shape of a croissant pastry.




\runinhead{MHD interpretation of cavity morphology: expanded, but trapped, twisted flux.}
Simulations have demonstrated that a flux rope expanding outwards into closed magnetic fields may find an equilibrium configuration as the forces causing the expansion are countered by confining magnetic tension forces \citep{gibfan_06b}.   Because magnetic field strength drops off with height, there is greater lateral confinement than vertical, resulting in an equilibrium flux rope that is taller than it is wide. The equilibrium flux rope will then have an arched, tunnel-like morphology with narrow aspect ratio (Figure \ref{fig:croissant} (right)).   

\begin{svgraybox}
\runinhead{Open Questions}
\begin{itemize}
\item Do larger cavities observed in white light have the same 3D morphology as measured for smaller, EUV cavities? 

\end{itemize}
\end{svgraybox}

\subsection{Thermal properties}
\label{subsec:thermobs}

Once the three-dimensional morphology of a cavity is established, it is possible to quantify density and temperature and so gain insight into its thermal properties.

\begin{figure}[b!]
\center{\includegraphics[scale=.34]{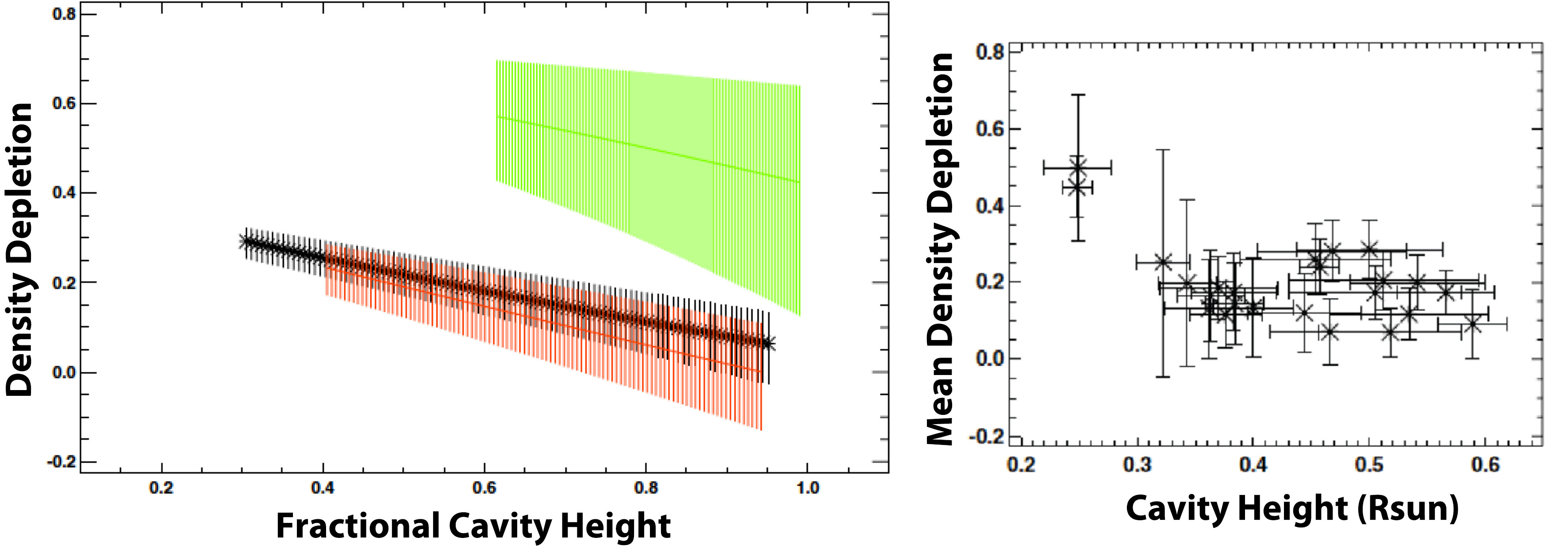}
}
\caption{Fractional density depletion of cavity relative to surrounding streamer  self-similarity.  Left: normalized radial profiles of depletion for three cavities. Right: average depletion for 24 cavities.  Green profile (left) corresponds to one of the small outlier cavities (right) (from \citet{fuller_09} - reproduced by permission of the AAS.)} 
\label{fig:Depletion}       
\end{figure}

\subsubsection{Density}
\label{subsubsec:dens}

Cavities observed simultaneously in multiple wavelengths provide strong evidence that they represent a depletion of coronal plasma density \citep{gibson_10}.  Perhaps the most straightforward density analysis utilizes white light data, which has no temperature sensitivity.  An analysis of 24 white-light cavities \citep{fuller_09} used geometric arguments and Van de Hulst \citep{vandehulst} inversion to measure density within cavities and surrounding streamers.  This study found that the average density depletion (relative to surrounding bright coronal streamers) as measured at the center of the cavity and just above the occulter ($1.16 \Rsun$) was $25\%$, and the maximum depletion was $60\%$.   An analysis using EUV line ratios (to reduce temperature dependency) in conjunction with white light data and explicitly accounting for 3D cavity morphology was able to measure density at heights above the limb as low in height as $1.08 \Rsun$, where it was found to be $30\%$ depleted relative to a surrounding streamer \citep{schmit_11}.  These measurements are consistent with previous analyses using radio data \citep{straka,kundu,marque_04}.  Thus, cavities are by no means empty, and are significantly brighter than coronal holes at similar heights.

Another interesting aspect of cavity density structure, at least for white light cavities, is that the density depletion of the cavity relative to the surrounding streamer is generally maximum at low heights, and minimum (indeed, often zero) at the cavity top \citep{fuller_08,fuller_09} (see \citet{gibson_10} for a discussion of the implications for hydrostatic scale heights).   When cavities are normalized to their top height, a curious self-similarity is apparent.  All cavities studied have similar depletion slope, independent of size or time of solar cycle (Figure \ref{fig:Depletion} -- left).  Moreover, the mean depletion is essentially the same for the majority of cavities, with the exception of two small outlier cavities (Figure \ref{fig:Depletion} -- right) which possess greater depletion at all heights, even at the cavity tops.  Whether such a jump in density at the top of cavities is typical for small cavities, which are likely to be under-selected in any white-light study, remains to be demonstrated through a quantitative survey of emission cavity depletion.

\runinhead{MHD interpretation of density depletion: enhanced magnetic pressure, field-line-length dependence, and/or stability selection effect?}
Low density within a magnetic flux rope might be expected if there is a jump in magnetic field strength across its boundary, for example due to an enhanced axial field relative to surrounding fields, and with it a current sheet or layer.
Then, total pressure continuity would require a decrease in thermal pressure within the cavity to balance the increase in magnetic pressure, and if the temperature is essentially the same in the cavity and surrounding streamer (see Section \ref{subsubsec:temp}) this would mean a decreased density.

Such continuity arguments are most relevant at the flux rope boundary, however, and depending upon the rope's twist profile there may be little (or no) jump in axial field at this boundary.  Indeed, simulations of flux ropes emerging into a potential field (e.g. \citet{gibfan_06b}) result in a near-force-free equilibrium with a smoothly varying boundary. It is true that current sheets do form at the boundary of these simulated ropes when they are dynamically perturbed, at least at lower heights, and it only takes a small jump in magnetic pressure to require a relatively large jump in thermal pressure in the magnetically-dominated (low plasma $\beta$) corona.  This may explain why the elliptical boundary of the cavity is generally so sharply defined (at least at lower heights).  But why is the center of the cavity persistently depleted in density?

\begin{figure}[t!]
\center{\includegraphics[scale=.235]{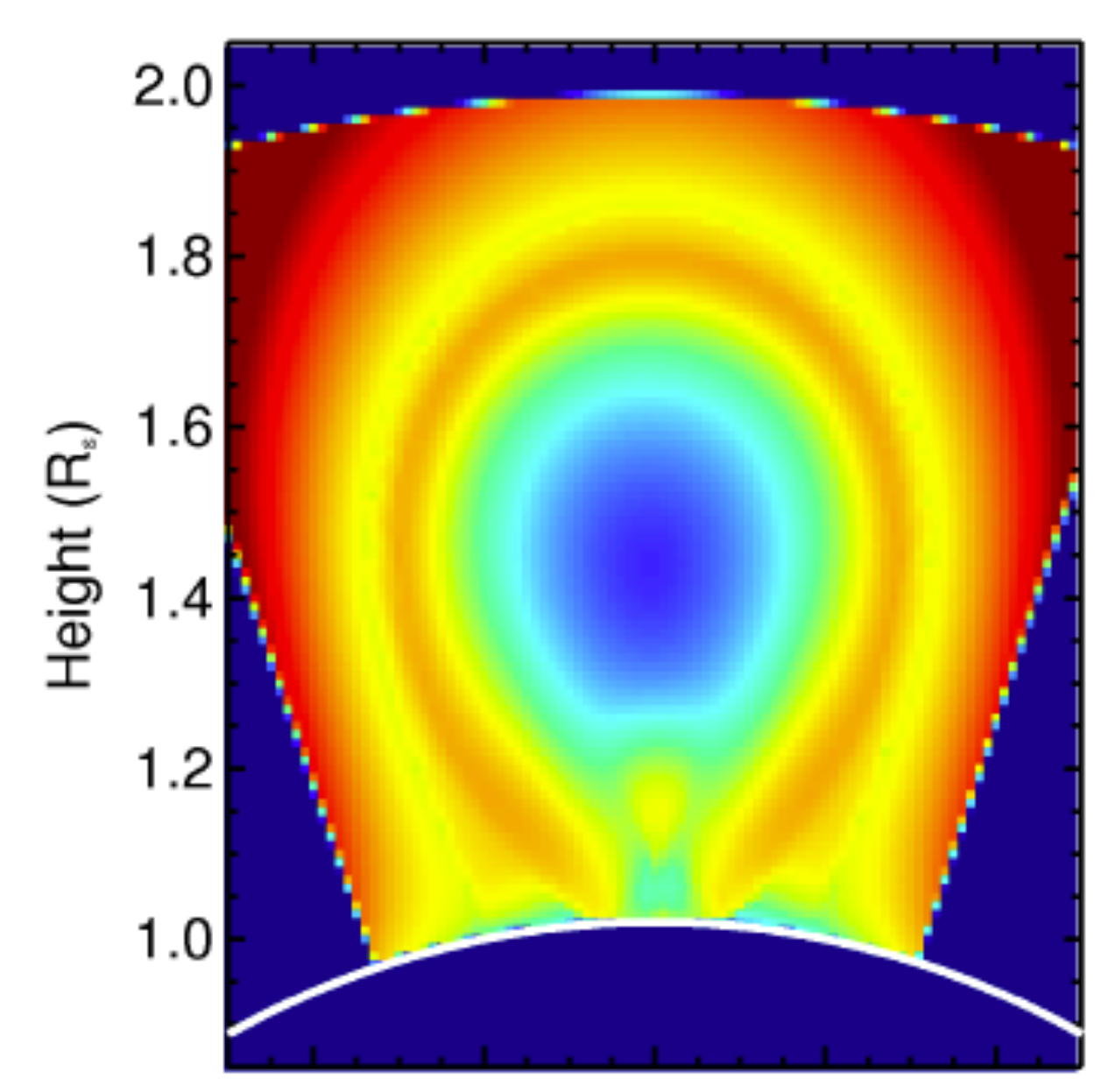}
\includegraphics[scale=.235]{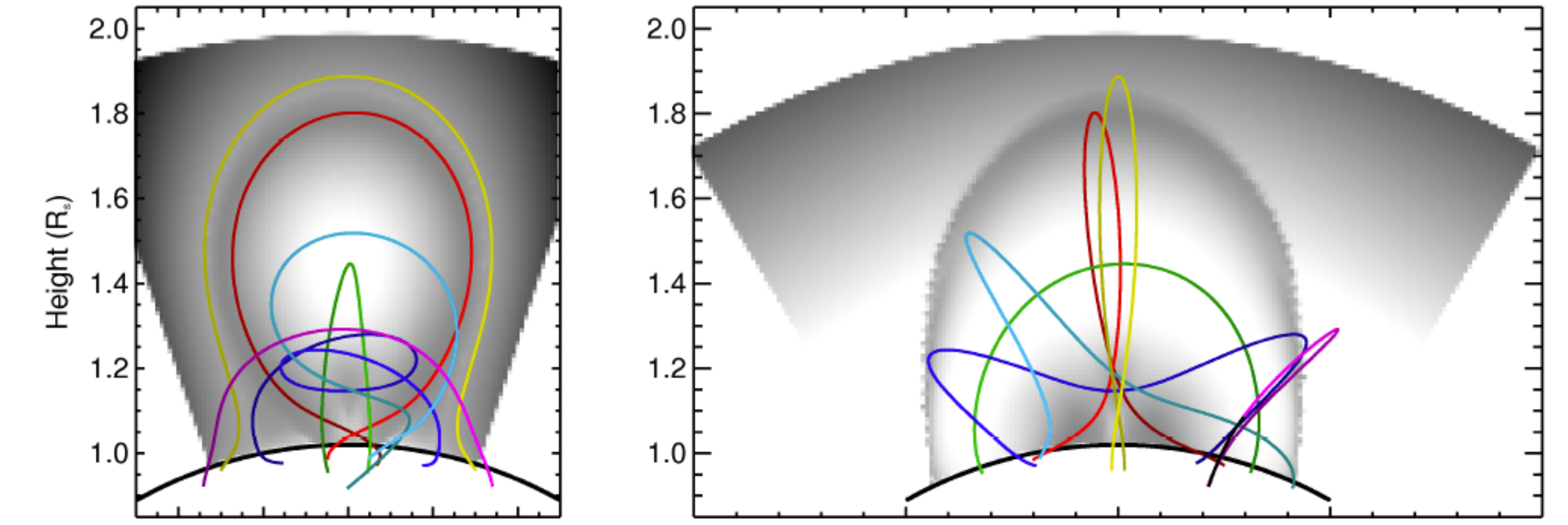}}
\caption{Arched flux-rope model illustrating how axial field lines may be non-dipped and relatively short.  Left: color contours representing length of field lines intersecting a plane (blue=short, red=long) (from \citet{schmit_14}); Middle/right: views along/normal to the LOS of sample field lines (Courtesy D. Schmit).  The central, axial lines are the shortest (e.g., green line).  A slight decrease in length can be seen in the left-hand plot at the boundary between the outermost rope lines (e.g., red line) and surrounding arcade (e.g., yellow line). Note that field lines wrapping around the outer part of the rope  are dipped (e.g., dark blue) but the central axial field lines (e.g., green) are not.} 
\label{fig:length}       
\end{figure}

It is clear that density will be controlled by thermodynamic variation within the rope.  This variation in turn has its origins in the magnetic structure, and may be affected for example by processes that depend upon magnetic field line length. Generally speaking, field line length increases from the axis of the rope outwards (Figure \ref{fig:length} -- left).
Arguments have been made that line-of-sight integrated density will increase from cavity center outwards if the magnetic structure is a flux rope, due to increased number of particles moving along the progressively longer, more twisted field lines towards the exterior of the flux rope, and assuming an ongoing flow throughout the cavity \citep{krallchen_05}.  
In fact, a flow is not generally required for a dependence of density on field line length:  solutions of hydrostatic equilibrium along field lines extracted from a simulated, stable flux rope (e.g., Figure \ref{fig:length} -- middle/right) indicate that the short (and low-lying) axial field lines would be depleted about 35\% relative to the outermost rope (and surrounding arcade) field lines at the same height.  This arises in part from a weak linear dependence between density $n$ and field line length $L$ ($n \sim Q^{(4/7)} L^{(1/7)}$) under the assumptions of constant footpoint temperature and uniform heating $Q$ along field lines. In addition, gravity requires that higher-lying field lines are supported with a higher pressure at their base, and consequently are more dense overall \citep{schmit_14}.
Such a hydrostatic description may only be appropriate for the short, axial field lines.  Thermal nonequilibrium (TNE)  (see \citet{karpen14_book}) is likely to be significant along the longer field lines of the outer portions of the rope and the surrounding arcade \citep{klimchuk_10}.  However, this would only emphasize the contrast with the short, axial field lines at rope center, since TNE would result in further density enhancement relative to a hydrostatic state \citep{schmit_13b}.

If field line length does affect density, the degree of depletion resulting will depend upon the twist profile of a flux rope, and on its size and shape relative to overlying unsheared fields.  The flux rope shown in Figure \ref{fig:length} possesses twist between one and two full turns, and has an overall morphology (height, length, radius) similar to that measured for a PCF cavity \citep{gibson_10}.  In such a geometry, the central, axial field lines are the shortest, and
indeed are significantly shorter than the outer flux rope and overlying arcade-like field lines. If the rope were less arched, the arcade field line length and height would be closer to that of the axial field line, and the depletion would be less.  If the rope were more twisted, the outermost flux rope lines would be even longer than the axial field lines, and the depletion could be greater.  Thus, measurements of the density depletion of cavities in combination with their 3D morphologies may constrain the possible twist profiles of the magnetic structures associated with them.

Finally, the characteristic density profiles of cavities -- similar mean depletions, an upper depletion  limit of about a factor of two, and decreasing depletion with height -- may be a consequence of a selection effect associated with their stability.  Magnetostatic solutions exist where the plasma in the flux rope has higher density than its surroundings, but these would be unstable both to magnetic curvature forces and to the Rayleigh-Taylor instability in their lower halves where heavy material lies above light \citep{lowhund}.  Underdense flux ropes (thus, cavities) are possible, since magnetic curvature would be stabilizing in such cases, but if their density becomes too low they could be become unstable to buoyancy \citep{low_82}.  A much more twisted flux rope could result in longer field lines at the rope boundary, and thus potentially a more deeply-depleted rope center, but such a configuration would be prone to the kink instability.  Thus, perhaps, if cavities were other than we measure them, they would not survive long enough to be measured.

\begin{svgraybox}
\runinhead{Open Questions}
\begin{itemize}
\item Is decreased density in cavities related to field line length, and if so, what are the dominant thermodynamic processes involved?
\item Is the upper limit on cavity depletion (about a factor of two) due to stability selection effects, and if so what are the relevant instabilities involved? 
\item Could the decrease in density depletion towards the top of white light cavities be a stability selection effect where Rayleigh-Taylor-unstable boundaries are avoided? 
\item Could the narrow (prolate) elliptical aspect ratio seen in small EUV cavities result in a more strongly stabilizing magnetic curvature force, and thus enable a greater degree of depletion at their tops?
\end{itemize}
\end{svgraybox}

\subsubsection{Temperature}
\label{subsubsec:temp}

Establishing the temperature of cavities requires a multiwavelength analysis, and ideally one which first takes into account both 3D morphology and density.  In such a study, the average cavity temperature was found to be essentially equivalent to that of the surrounding streamer \citep{kucera_12}.  However, the cavity exhibited more thermal variability than the streamer, indicating that multiple temperatures were present for a given height in the plane of the sky (note that contributions to the line-of-sight intensity integral were established to be dominated by plasma actually lying within the extended tunnel of the cavity).   Previous analyses have argued that cavities may be cooler than their surrounding streamer \citep{guhaetal_92} or hotter \citep{fuller_08,vasquez09,habbalcav}. Thus there does not yet seem to be a non-ambiguous answer to the question as to whether the low-density plasma within cavities is typically hotter or cooler than that in the surrounding streamers (although the evidence for at least some hot material in the vicinity of cavities is compelling -- see further discussion in Sections \ref{subsec:flowsub}-\ref{subsec:subobs}).  

\begin{svgraybox}
\runinhead{Open Questions}
\begin{itemize}
\item What is the temperature of the cavity relative to a surrounding streamer? 
\item Does it depend upon where in the cavity one looks?
\item What is the nature of multithermal variaton within the cavity?
\end{itemize}
\end{svgraybox}

\subsection{Flows}
\label{subsec:flowsub}

Two factors may complicate cavity temperature analyses: substructure and flows.  We will discuss substructure in Section \ref{subsec:subobs}, and in particular the hot cores, or ``chewy nougats'' apparent in SXR cavities.  First, however, we will discuss the flows, both in the plane of the sky (POS) and along the line of sight (LOS), which may contribute to the thermal variability within cavities.

\begin{figure}[h!]
\center{
\includegraphics[scale=.16]{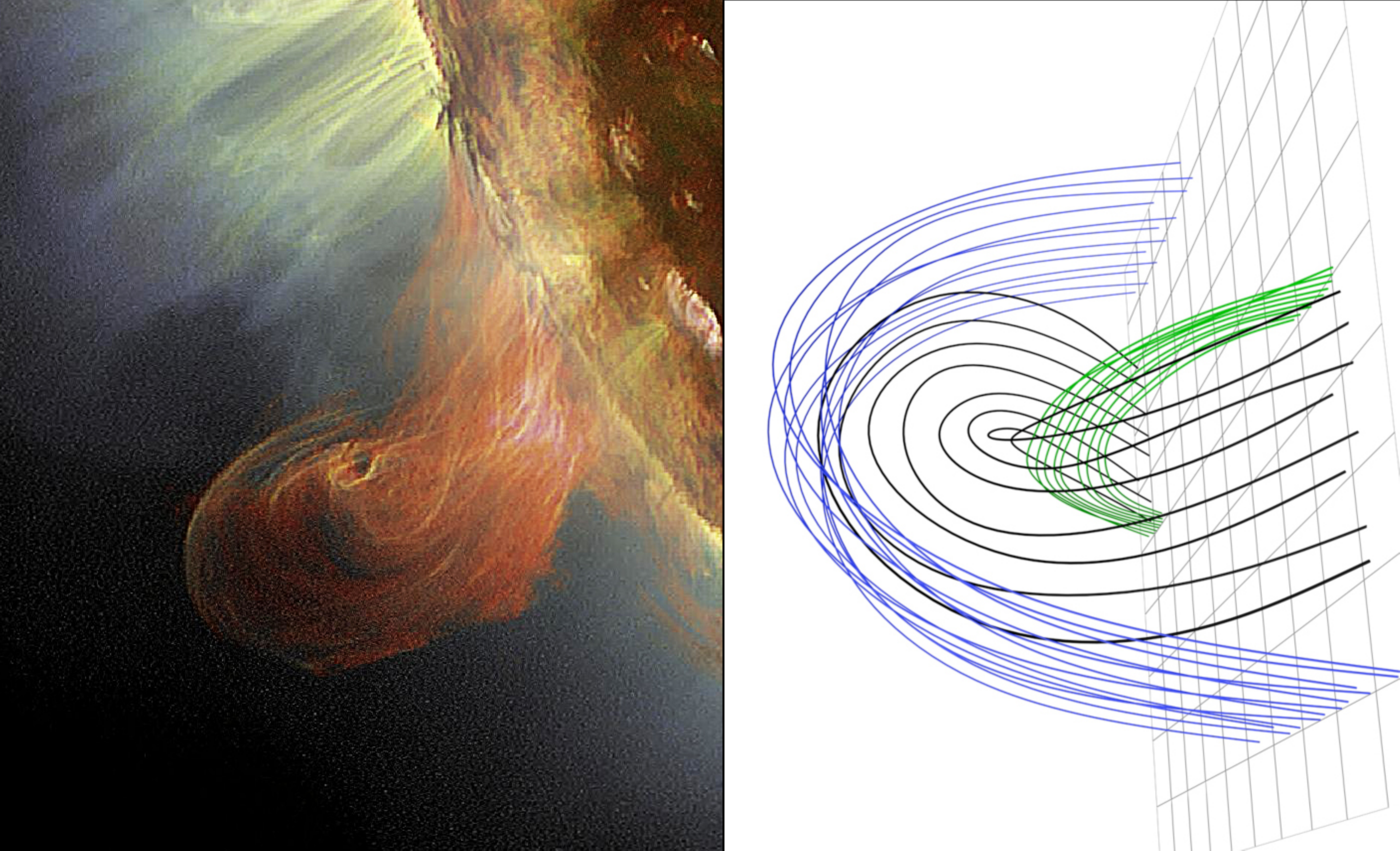}
\includegraphics[scale=.2]{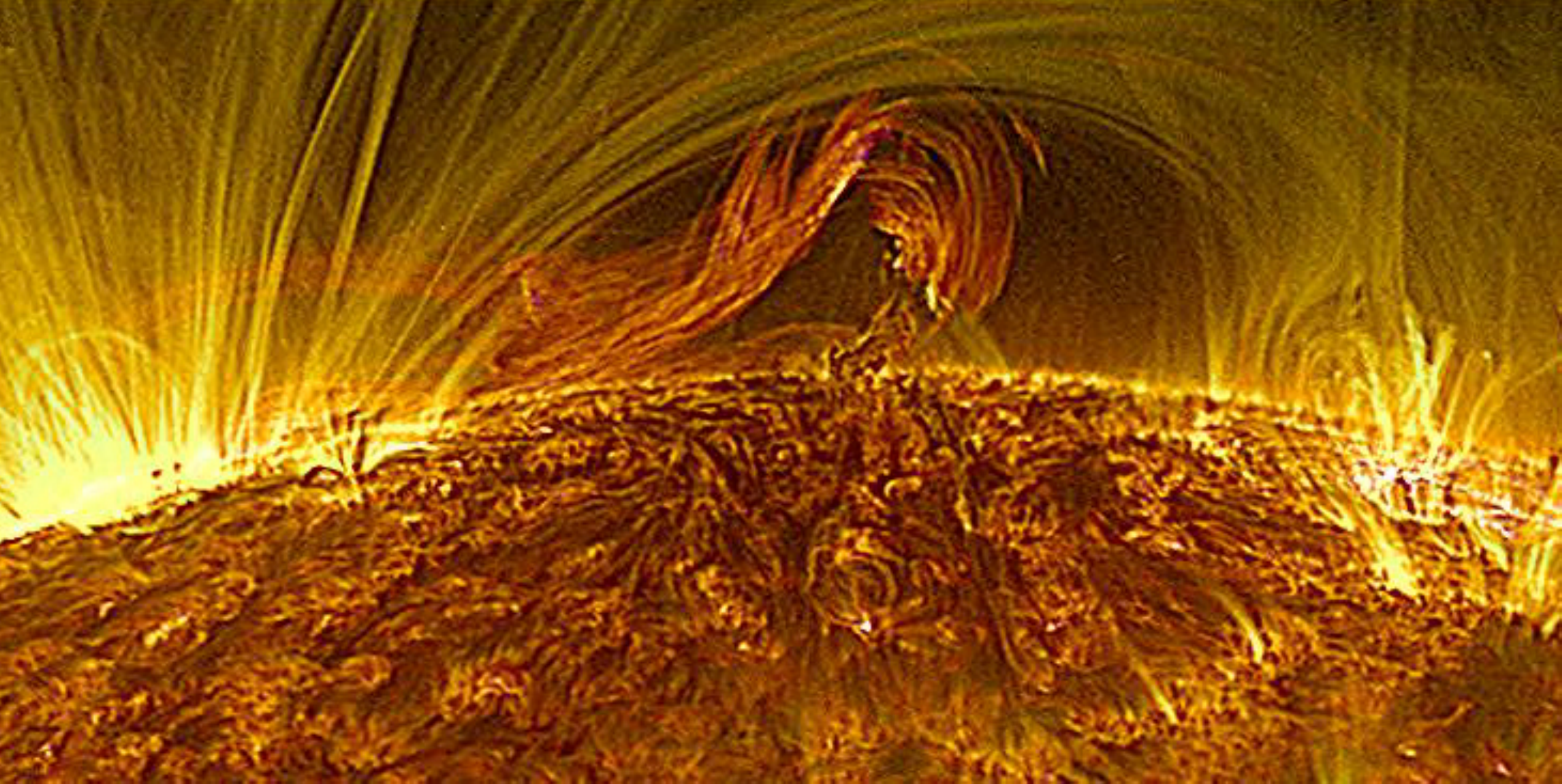}
\includegraphics[scale=.28]{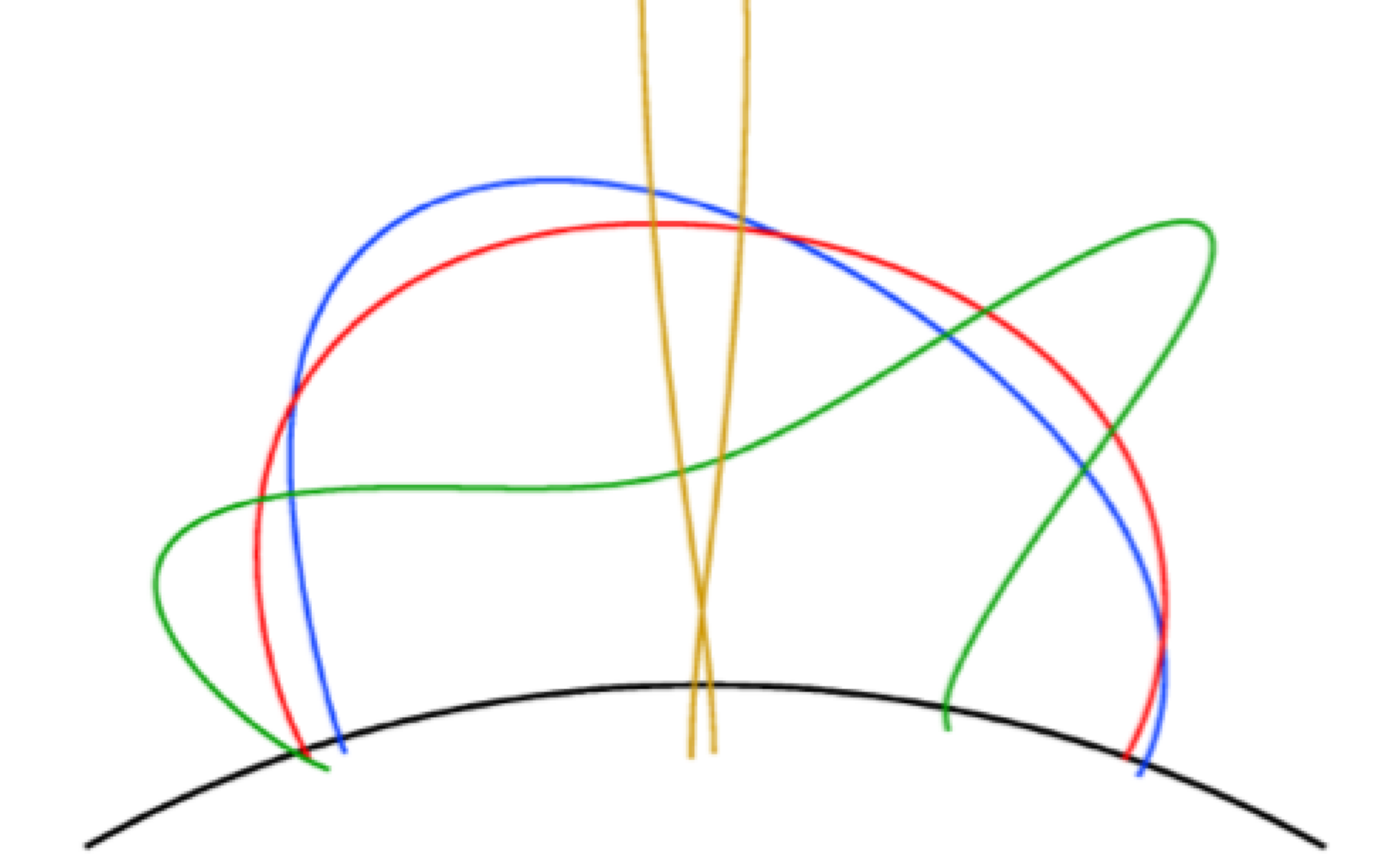}}
\caption{Plane-of-sky flows trace out writhing field lines.  Left: SDO/AIA multi-wavelength observations of apparent swirling motion within cavities (top: 16 June 2011, bottom: 23 July 2012 (from \citet{panasenco_13}; used by permission)).  Right: a non-flux-rope model (top) representation of magnetic field lines of filament spine (green), coronal loops (blue), and cavity (black) (from \citep{panasenco_13}; used by permission), and (bottom) sample field lines extracted from an MHD equilibrium flux rope model (described in \citet{gibfan_06b}) (adapted from \citet{schmit_14}. } 
\label{fig:posflows}       
\end{figure}

\begin{figure}[b!]
\center{
\includegraphics[scale=.35]{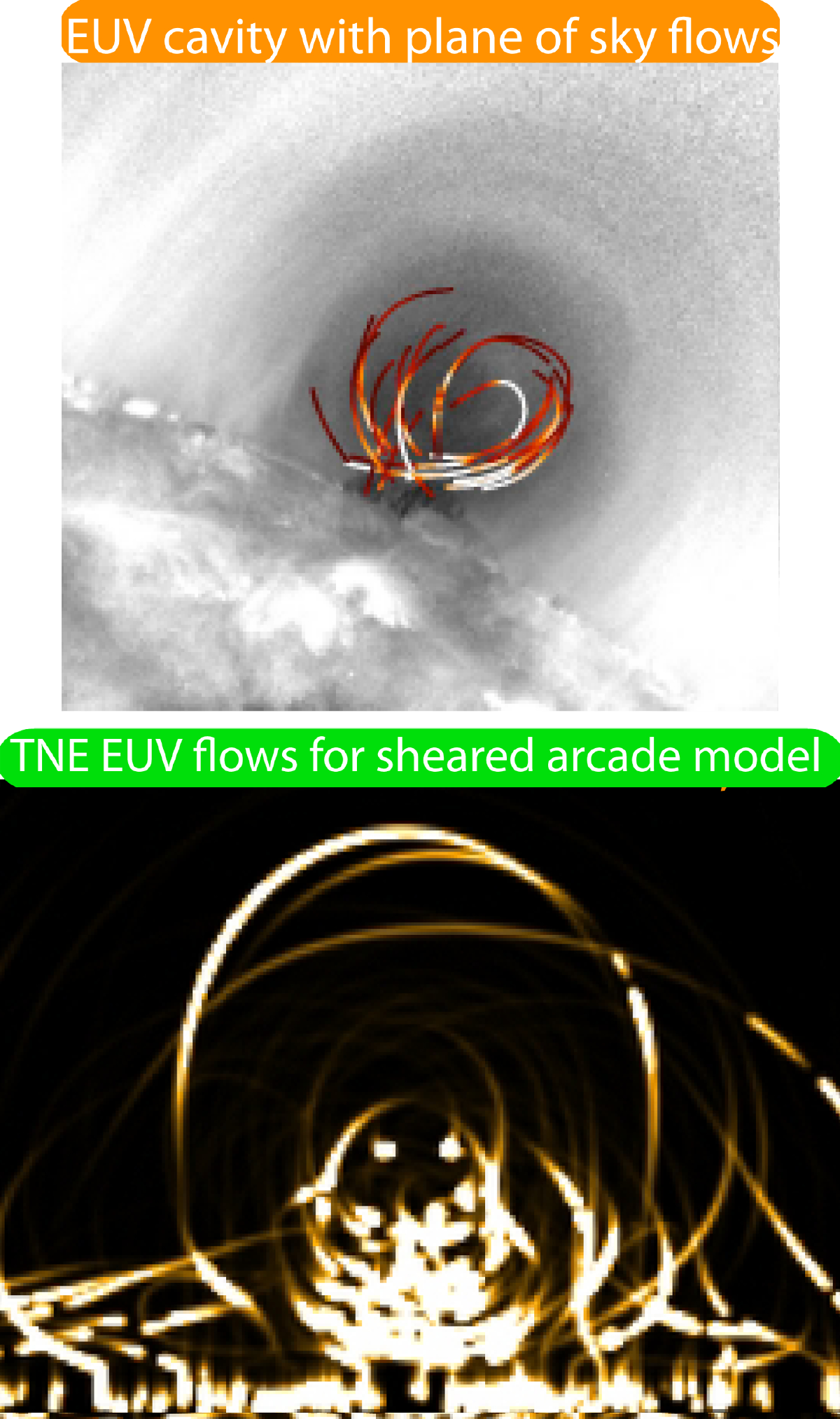}
\includegraphics[scale=.35]{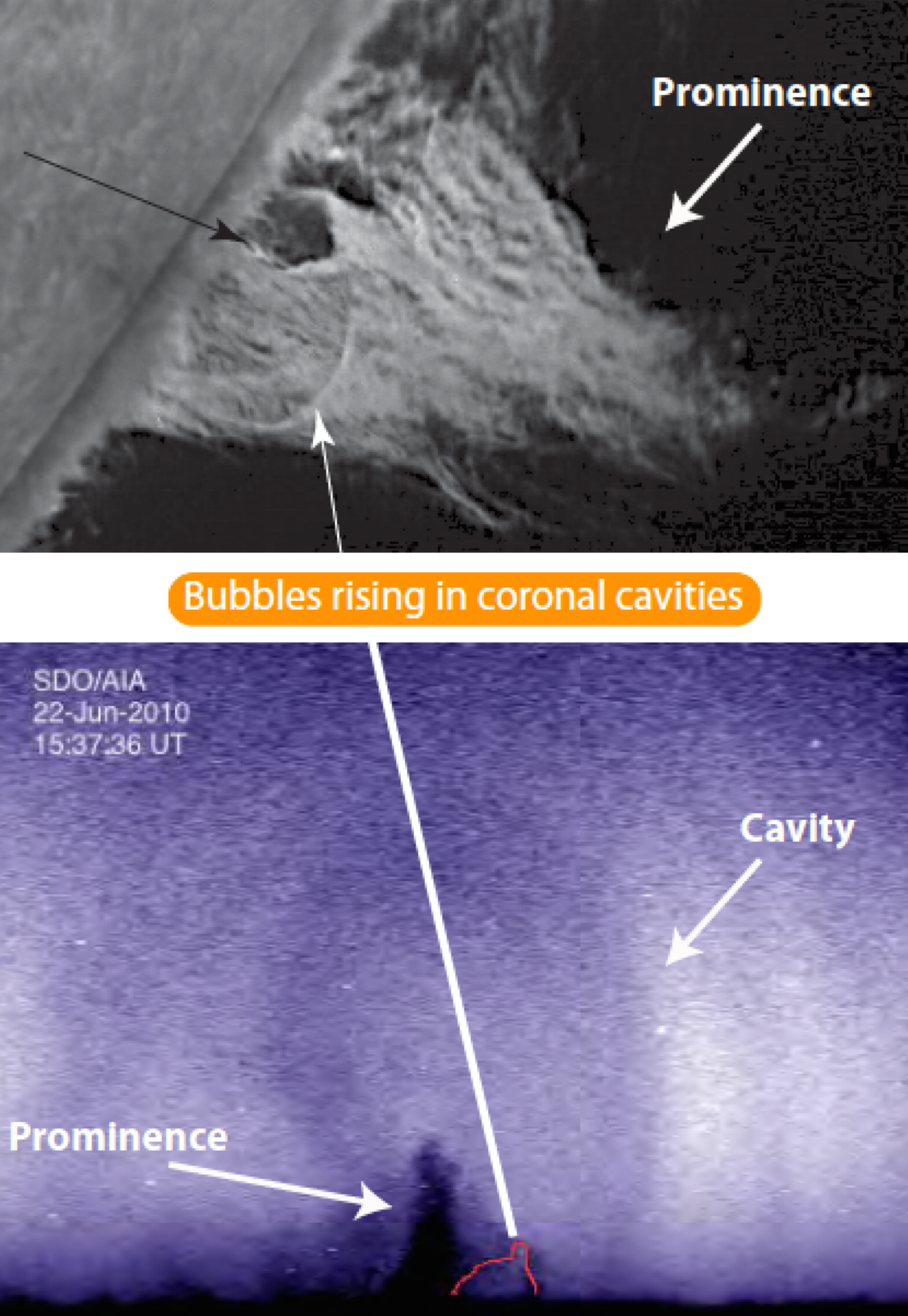}}
\caption{Plane-of-sky flows connecting prominences and cavities. Left: prominence-cavity interface flows along "horns"; (top) multiwavelength SDO/AIA observations, with paths of POS flows extracted (adapted from \citet{schmit_13a}); (bottom) synthesized flows driven by TNE in a sheared-arcade model (from \citet{Luna.Karpen.promin.thread.HD.2012ApJ...746...30L}; reproduced by permission of the AAS).  Right: Bubble rising through prominence (Hinode/SOT) into coronal cavity (SDO/AIA) (from \citet{BergerT.bubble-RT-instable.2010ApJ...716.1288B,bergerasp_2012}; reproduced by permission of the AAS).} 
\label{fig:hornbubble}       
\end{figure}

\subsubsection{POS flows}
\label{subsubsec:flowpos}

Even cavities in a quiescent, non-erupting state exhibit a range of interesting dynamics.   High-time-cadence movies of EUV images indicate the presence of apparent swirling motions in plane of sky projection within cavities \citep{li_12} (Figure \ref{fig:posflows}, left).  These flows have POS speeds in the range $5-10~km~s^{-1}$, and can persist in the same sense of rotation for several days \citep{wang_10}.     Another type of flow that has been associated with cavities involves plumes or bubbles (which may have sizes on the order of $10~Mm$) rising through the prominence and into the cavity (Figure \ref{fig:hornbubble}, right)
\citep{BergerT.promin-plume.2008ApJ...676L..89B,detoma_08}. The ascent speed of these bubbles is on the order of tens of $km~s^{-1}$ \citep{BergerT.bubble-RT-instable.2010ApJ...716.1288B,bergerasp_2012}, and simultaneous optical and EUV observations indicate temperatures that are $25-120$ times hotter than the overlying prominence \citep{Berger.hot-bubble.2011Natur.472..197B}.  Whether these represent temperatures within the bubbles and plumes themselves, or whether they are indicative of the background corona/cavity seen through gaps in the prominence opened up by the bubbles/plumes, remains a subject of debate \citep{dudik_12}. 

In addition, transient EUV brightenings within cavities occur in the form of "horns" in $171~\AA$  which extend nonradially from the top of prominence observed in $304~\AA$ (Figure \ref{fig:hornbubble}, left-top) \citep{schmit_13a}.   By analyzing a database made up of 48 such horns, spatial and temporal correlations were found between the coronal flows within the cavity, and flows associated with cooler, prominence plasma: in general, the formation of horns in $171~\AA$ preceded the formation of essentially co-spatial prominence extensions in $304~\AA$ by about 30 minutes.  

Variability  in EUV, and especially at $171~\AA$, is much greater in cavities than in surrounding streamers.  It may be that the multithermal nature of cavities arises from a dynamic prominence-cavity interface, which would project both cold and hot plasma into the main volume of cavity  \citep{schmit_13a}.

\runinhead{MHD interpretation of POS flows: tracing magnetic field lines and transferring helicity and mass.}

The thermodynamics driving horn-like flows between prominence and cavity may arise from thermal nonequilibrium  \citep{karpen14_book}, and Figure \ref{fig:hornbubble} (left) illustrates how a model of TNE-driven flows along a sheared-arcade magnetic configuration produces synthetic EUV images similar to observations.
Likewise, a model utilizing TNE-driven flows along field lines of a flux rope found that the temporal and spatial correlations between prominence and cavity flows were well captured, and in particular that EUV brightenings (horns) would be followed by prominence formation \citep{schmit_13b}.  As discussed in Section \ref{subsubsec:dens}, TNE would be most likely to occur on longer, dipped field lines; for the flux rope these lie along the outer boundary (Figure \ref{fig:length}).   Thus,  the multithermal, variable, density-enhanced part of the magnetic structure would actually only fill a small fraction of a flux rope volume, but would project into some or all of the cavity at the limb. 

In a low-$\beta$ regime such as the corona, coronal plasma must flow along magnetic field lines \citep{ballester14_book}. With this in mind, the swirling motions within cavities are certainly suggestive of flow along twisted magnetic field lines.  One must be very cautious of projection effects, however.  Indeed, analysis of these motions using multiple viewing angles implies that the motion is along strongly writhed field lines, but probably not ones that wind more than one full turn \citep{panasenco_13}.  The question remains, though, whether such a strongly writhed field line is part of a flux rope or not. As Figure \ref{fig:posflows} (right) shows, both weakly-twisted-flux-rope and sheared-arcade models involve writhed field lines of very similar geometric appearance.  

\begin{figure}[b!]
\center{\includegraphics[scale=.5]{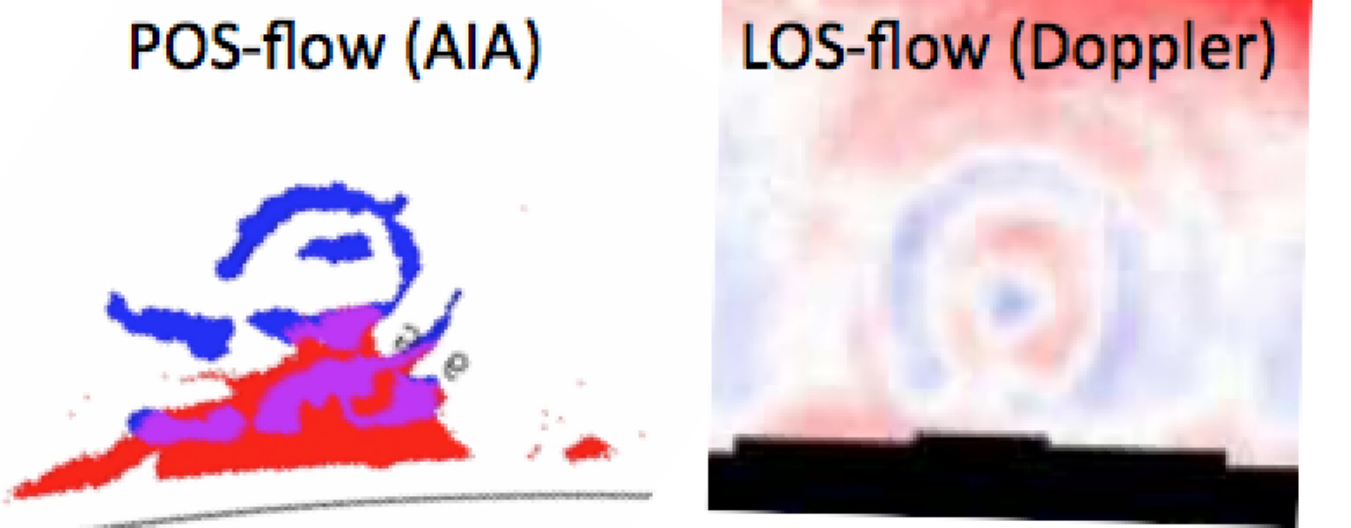}
\includegraphics[scale=.2]{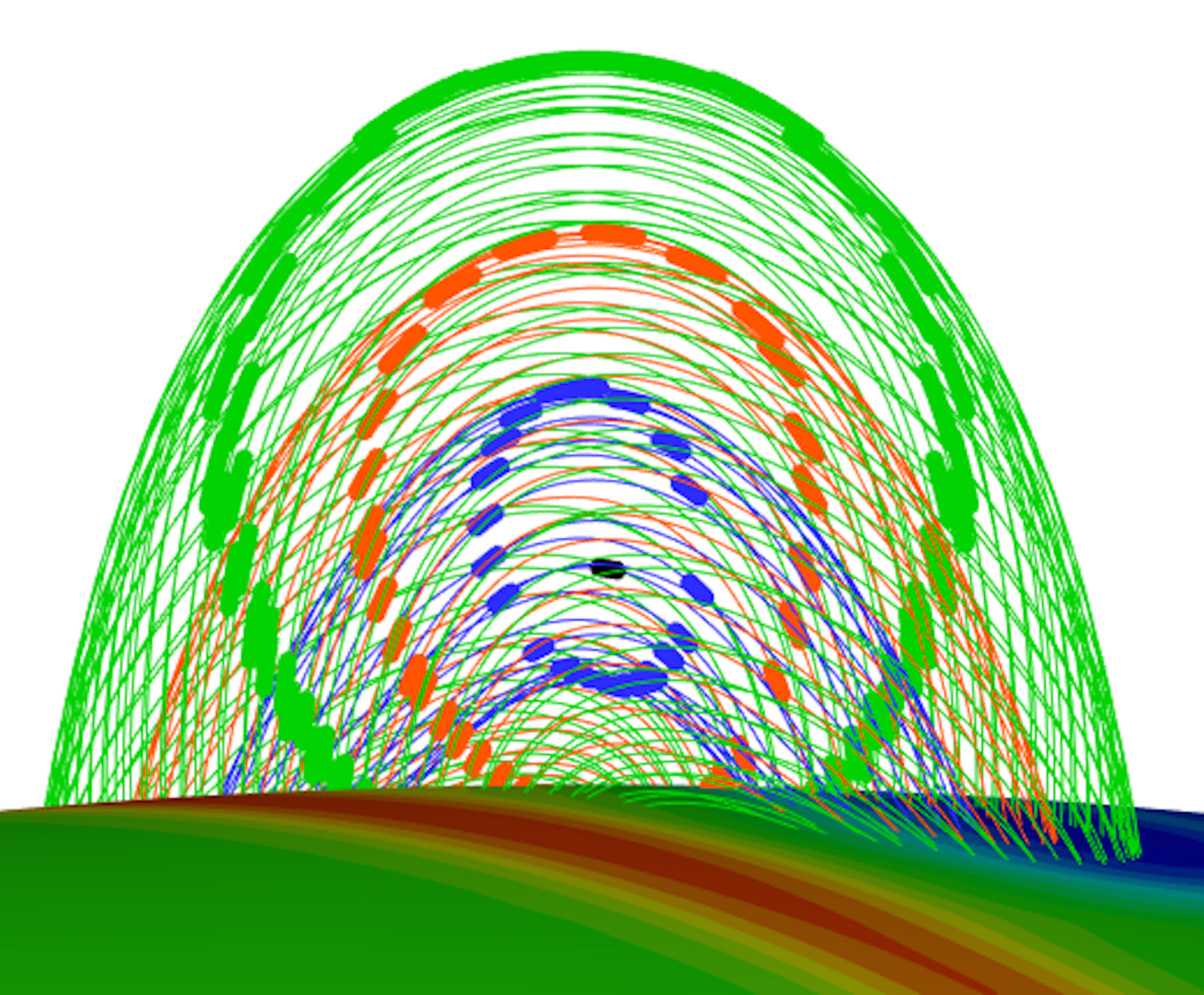}}
\caption{LOS 
and POS 
flows tracing nested rings about a central axis are strong evidence for toroidal flux surfaces.  
Left: Snapshot of POS flows along prominence-cavity interface horns, extracted from observations of SDO/AIA (blue -- $171~\AA$, red -- $304~\AA$, purple -- both) analyzed in  \citep{schmit_13a} (figure courtesy D. Schmit).  Middle:  LOS flows measured from Doppler shift by MLSO/CoMP (adapted from \citet{ula_13}).  Right:  flux surfaces of flux-rope model (described in \citet{fan_12}) shown by colored field lines, with intersection with POS indicated by bold-face colors.} 
\label{fig:fluxsurfaces}       
\end{figure}

The fundamental topological difference between these models is whether the writhed field lines {\it wrap around an axis}.  Since EUV flows light up only a portion of the magnetic field at any given moment, such observations will not generally capture both wrapping and axial lines
(although see Figure \ref{fig:fluxsurfaces} (left) for a possible snapshot of just such a configuration).  
For this reason, the field-aligned flows seen in POS projection, both swirling motions and flows associated with prominence-cavity horns, are unlikely to definitively distinguish flux-rope vs. sheared-arcade models.

The rising bubbles and plumes, in conjunction with prominence condensations and other downflows within the cavity, have been proposed together as evidence of a magneto-thermal convective mass cycle in the prominence and cavity
\citep{LiuW.Berger.Low.flmt-condense.2012ApJ...745L..21L, Berger.Liu.cavity.condense.2012ApJ...758L..37B}.  Although previous estimates have argued that there is not enough mass in the corona to account for the mass in the prominence \citep{saitotand_73}, such arguments assume a static corona, where in fact there is likely to be a  continuous input of plasma to the corona through TNE-driven flows, rising bubbles, spicules \citep{depontieu_11}, and reconnection-driven flows \citep{panasenco_13}.     Indeed, observations indicate that the total mass of a prominence at any given moment may be a small fraction of the mass condensing and draining through it \citep{LiuW.Berger.Low.flmt-condense.2012ApJ...745L..21L}.  

If the bubbles seen rising through the prominence and into the cavity are associated with magnetic flux concentrations \citep{dudik_12}, it may be that they also serve to transport small-scale magnetic twist upwards, merging with and adding to the helicity of the larger-scale fields of quiescent prominence cavities
\citep{Berger.hot-bubble.2011Natur.472..197B,low_12a,low_12b}.  Such a process could contribute to its ultimate eruption by leading up to an ideal instability or more generally to a state in which there no longer exists a stable minimum energy force-free equilibrium consistent with the increased helicity (see \citet{fan14_book} and further discussion below in Section \ref{subsubsec:teardrop}).

\begin{svgraybox}
\runinhead{Open Questions}
\begin{itemize}
\item Are rising bubbles within cavities associated with magnetic flux concentrations?
\end{itemize}
\end{svgraybox}

\subsubsection{LOS flows}
\label{subsubsec:flowlos}

Observations of Doppler line shift allow analysis of flows along the line of sight within cavities.  Such flows are common, with magnitudes of $5-10~km~s^{-1}$, length scales of tens of megameters, and durations of at least one hour \citep{schmit_09}.  The flows are spatially coherent, with boundaries that correspond to those of the cavity or to a central substructure.  Occasionally flows occur in the form of nested ring-like structures \citep{ula_13} (Figure \ref{fig:fluxsurfaces}: middle).  Interestingly, these nested flows may appear to be counterstreaming (alternating towards and away the viewer with radius).  The Doppler measurements shown in Figure \ref{fig:fluxsurfaces} incorporate a subtraction of the background coronal rotation using a median filter \citep{tian_13}, so there is some ambiguity about precisely where flows shift from towards to away in the rotating frame.  Nevertheless, there are clearly strong gradients in the flow of coronal plasma within the cavity, with functional dependence on radial distance from cavity center.

\runinhead{MHD interpretation of LOS flows: toroidal flux surfaces within a flux rope.}
If there is a ``smoking gun'' observation that indicates that the cavity is a magnetic flux rope, it is probably the nested rings observed in LOS flows.  When we see a coronal loop, the interpretation is that we are seeing a field line or collection of field lines along which plasma is being highlighted.  In the same way, when we see nested rings of LOS flows, or other ring-like or disk-like substructure within cavities (see Section \ref{subsec:subobs}), the implication is that the underlying magnetic structure of the cavity is one in which field lines trace out concentric rings when seen in projection at the limb.  
Figure \ref{fig:fluxsurfaces} shows magnetic field lines along flux surfaces, i.e., boundaries tangential to the magnetic field.   It is a property {\it distinct to the magnetic flux rope} that its flux surfaces are nested tori, and, when centered on the POS and with axis oriented along the viewer's line of sight, the intersection of these surfaces at the limb is one of nested rings (Figure \ref{fig:fluxsurfaces}: right).

Given a reasonably long, straight flux rope oriented along the LOS and centered on the POS, it follows that the LOS-directed component of a uniform flow will manifest as ring-like contours with a maximum at rope center, where the vector field and field-aligned flow is completely along the LOS. The fact that we do not see nested rings in every cavity must in part be due to the dependence of their observability on the orientation of the rope axis to the LOS.  The cavity has a finite length, and if it is not centered on the limb the asymmetries between foreground and background can eliminate the projection of flux surfaces as rings \citep{gibsoniau_14}.  A large curvature of the rope axis, or angle of the axis relative to the LOS, would similarly smear the rings out. 

It is less clear how the strong gradients in LOS flows arise between these rings.  One possibility is that plasma-emission weighting plays a role.  This could occur if a central substructure were at a temperature different enough from the rest of the cavity (see \ref{subsubsec:nougobs}) such that it had decreased emission in the wavelengths of light where the Doppler observations were measured.  Since the LOS flow measured is the integral over the line of sight, measurements at the center of the cavity would then be biased toward contributions away from the plane of the sky which, due to curvature in the rope, would sample plasma on field lines less aligned with the LOS and so contributing a reduced LOS flow.  Under conditions of constant flow, the maximum LOS velocity (integrated along the line of sight) would no longer be at the center of the cavity, but would lie along a ring at some distance from that center.

However, this argument cannot explain counter-streaming flows in nested rings.  For this, one requires a flow with a dependence upon radius within the rope, or equivalently that varies as a function of flux surface.  If the flux rope consisted of field lines that were uninterrupted in their wrapping around nested toroidal flux surfaces (thus, ergodic), different flows for each flux surface would indeed be expected. Because the flux rope is anchored in the photosphere, however, each flux surface consists of multiple field lines (set of colored lines in Figure \ref{fig:fluxsurfaces}).  Therefore, to create nested rings with counter-flows, one would have to drive the same flow along all field lines of a particular flux surface, but a different flow along all field lines of its neighboring flux surface.  If the flow is driven by differential heating at the fieldline lower boundaries, it is difficult to see why this would occur.  On the other hand,  it is possible that reconnection-driven flows might originate in the corona at a flux surface that represents a topological boundary \citep{fan_12} (see Section \ref{subsubsec:nougobs}).

\begin{svgraybox}
\runinhead{Open Questions}
\begin{itemize}
\item What drives the LOS flows seen within cavities? 
\item Do nested rings arise from flow excited coherently along flux surfaces, or does emission-weighting play a role?
\item Are the flows truly counter-streaming (relative to the LOS) between rings?
\end{itemize}
\end{svgraybox}

\subsection{Substructure}
\label{subsec:subobs}

\begin{figure}[t!]
\center{\includegraphics[scale=.45]{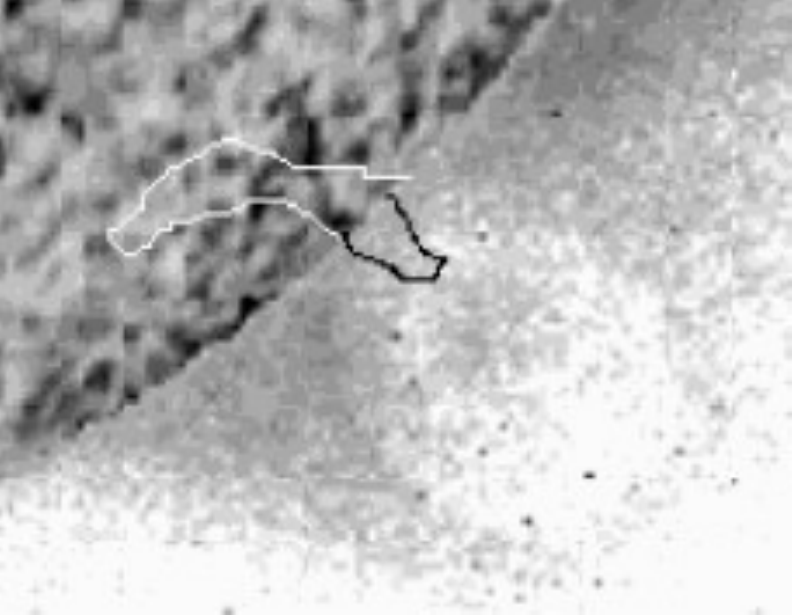}
\includegraphics[scale=.15]{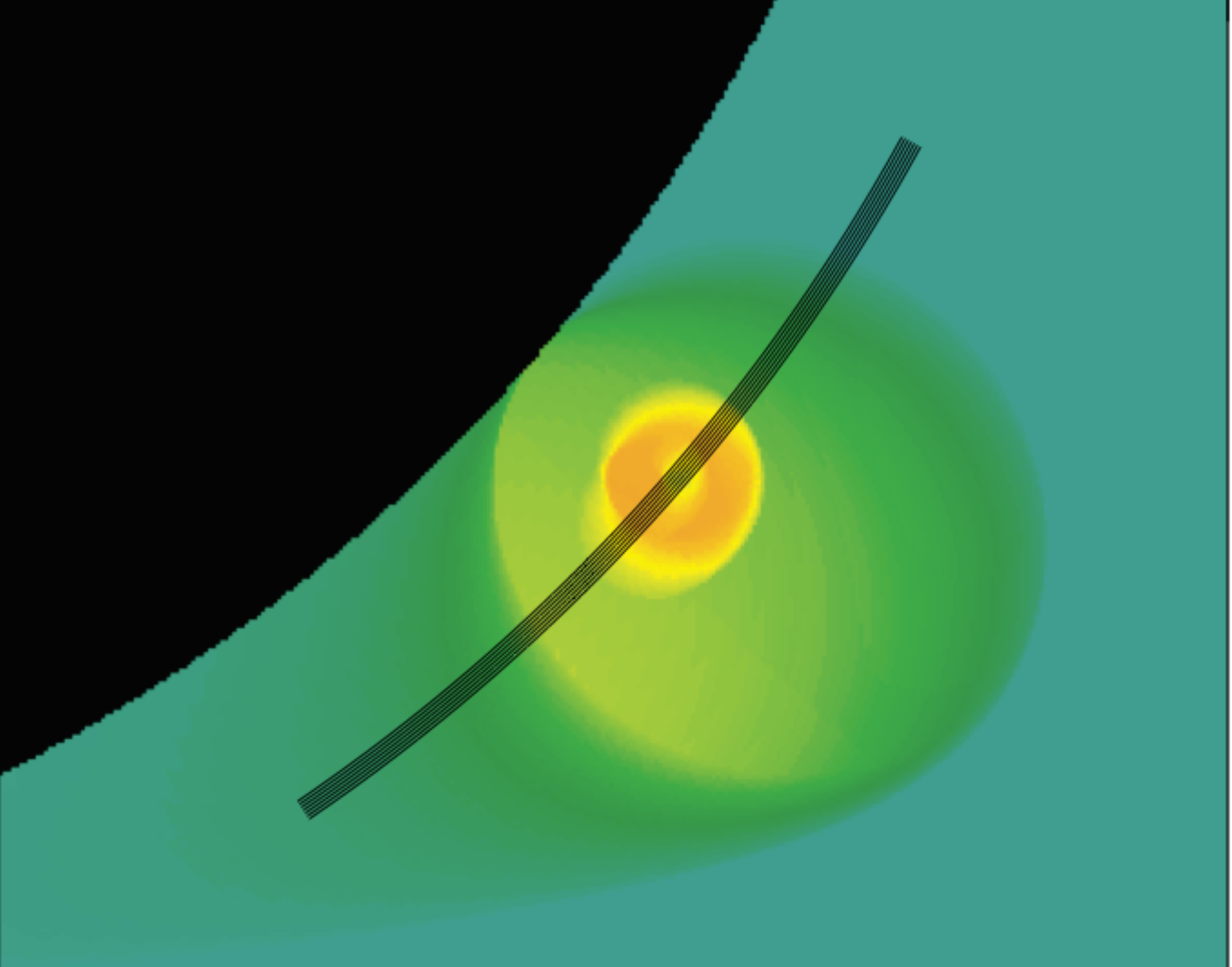}
\includegraphics[scale=.13]{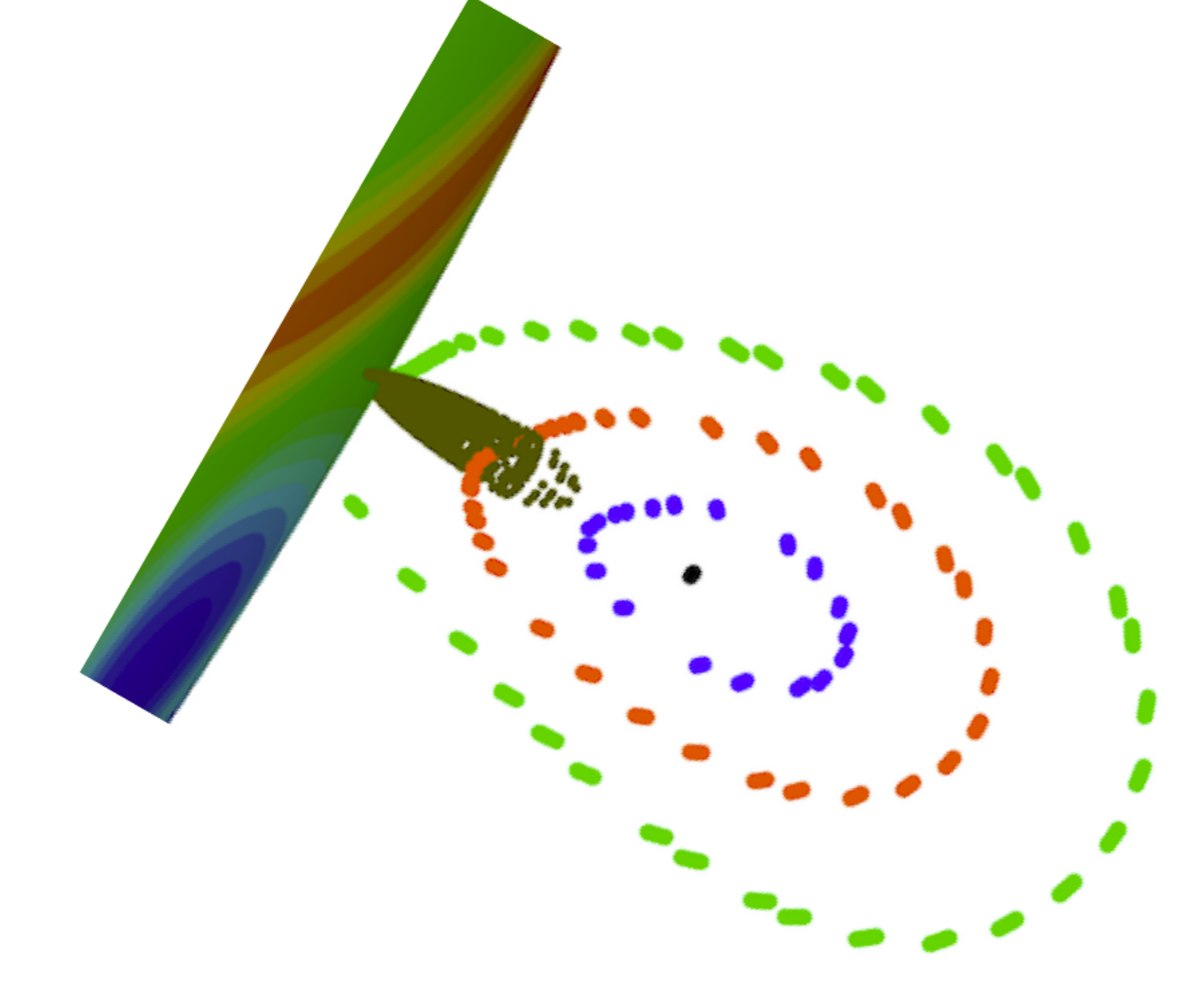}
\includegraphics[scale=.32]{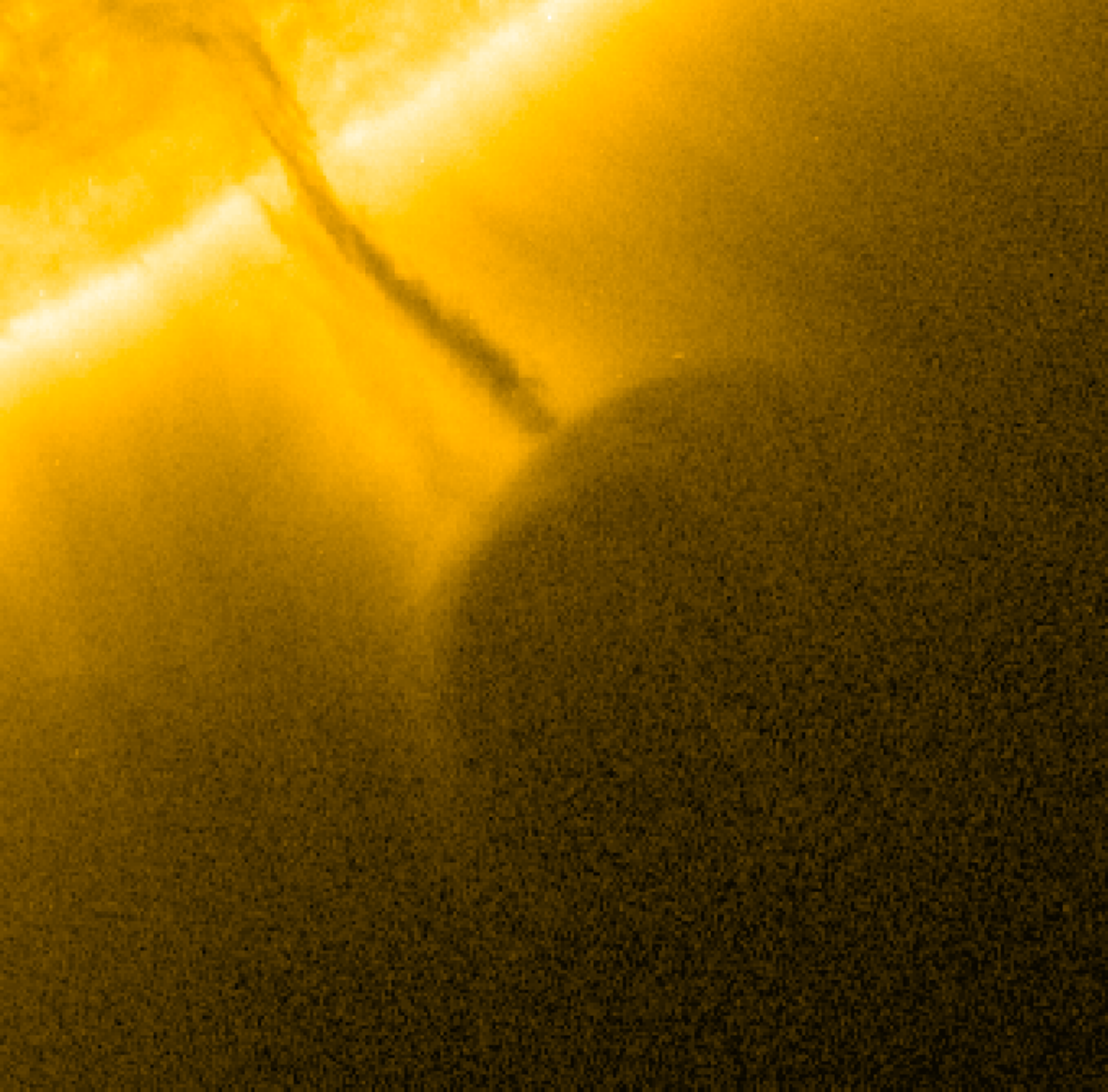}
\includegraphics[scale=.32]{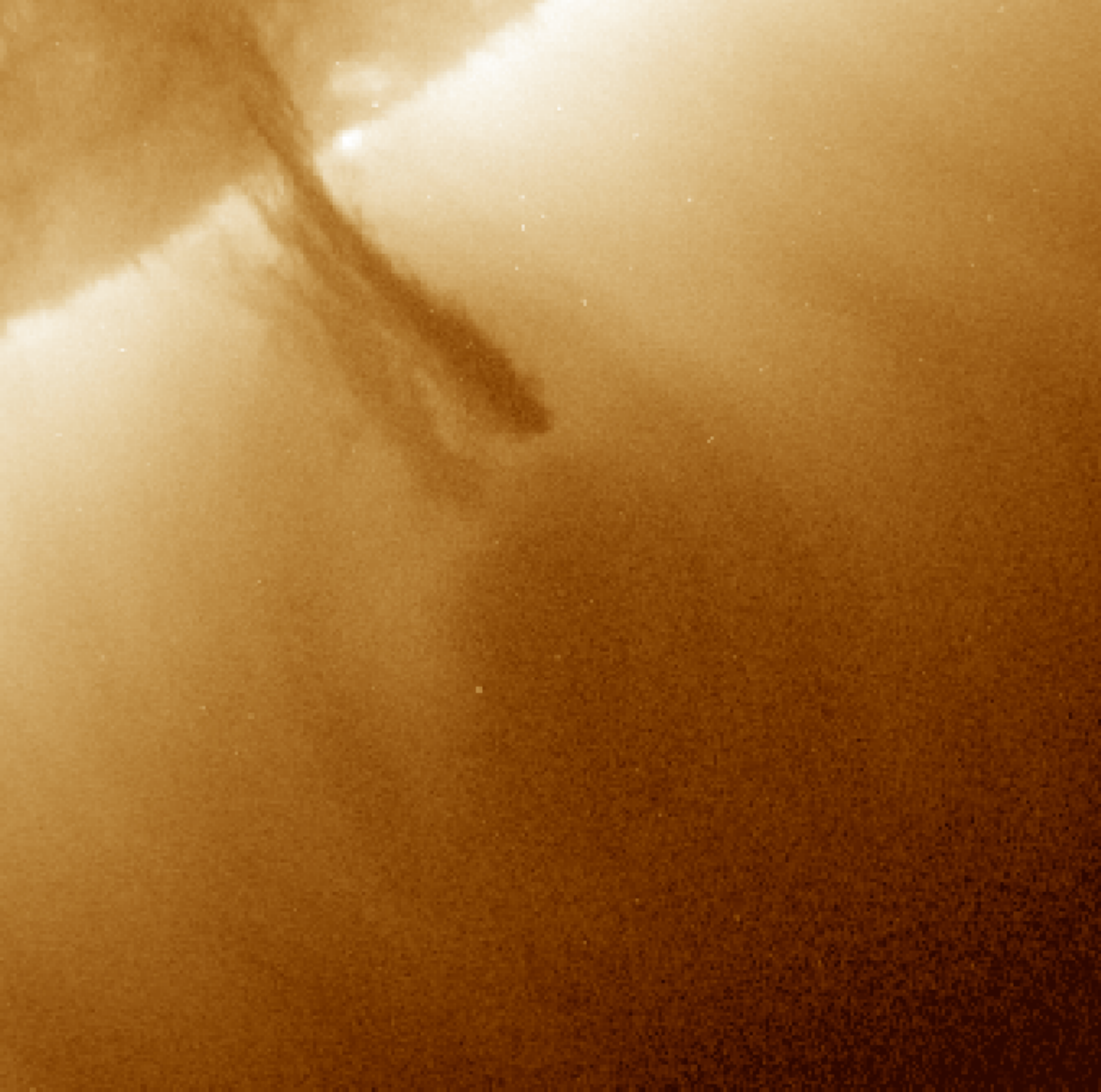}
\includegraphics[scale=.32]{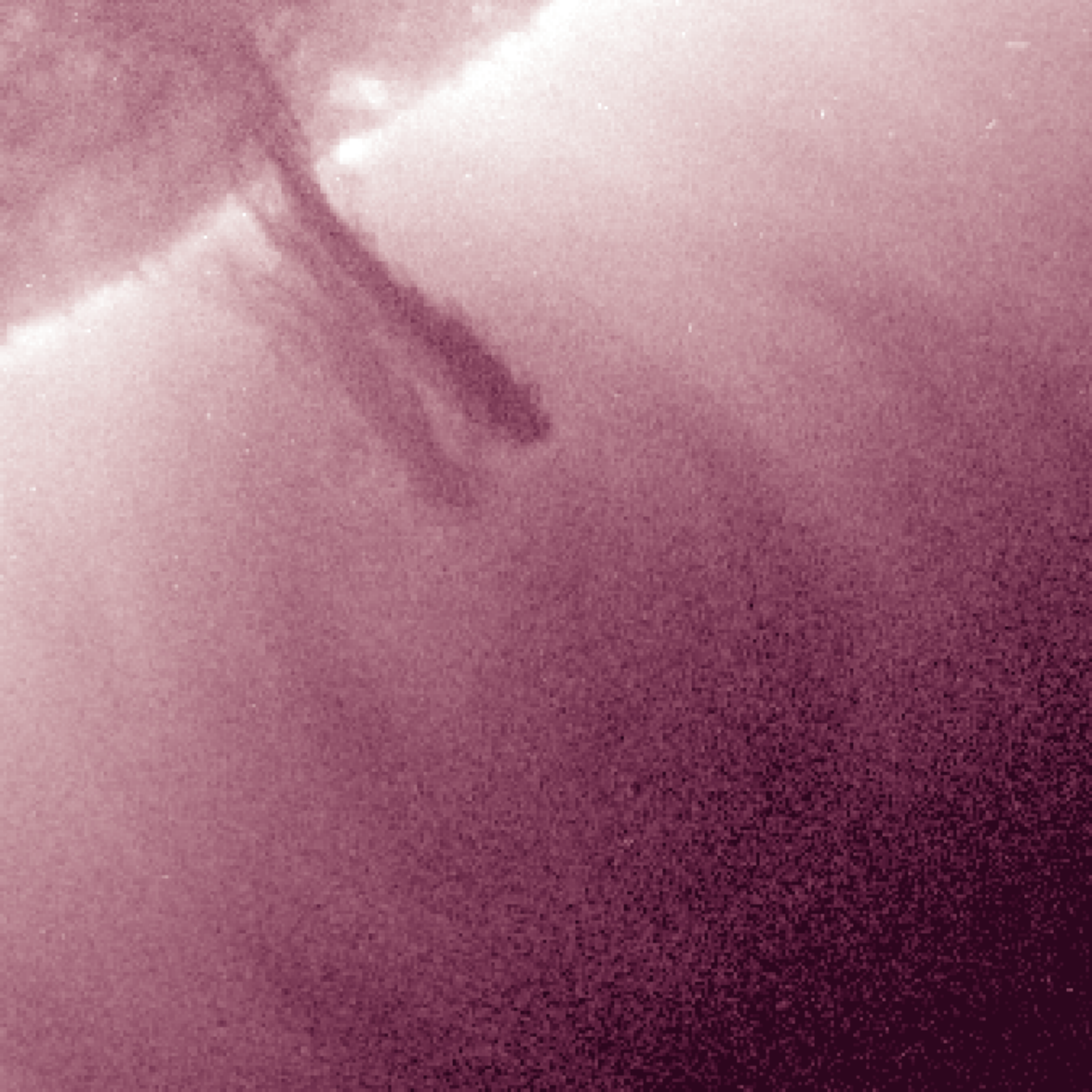}
}
\caption{Hot cores and central voids exist within larger cavity volume, perched atop prominences. Top left: Yohkoh soft-Xray observation of long-lived nougat (inverse image); location of H$\alpha$ prominence indicated by contour (from \citet{hudson_99}; reproduced by permission of the AAS).  Top middle: Temperature fit to multi-filter Hinode/XRT observations of nougat within cavity (adapted from \citet{reeves_12}). Bottom row: SDO/AIA images of teardrop-shaped cavity with dark disk-like void above prominence (so-called ``UFO'' cavity of March 11-12, 2012); left to right, 171, 193, and 211 $\AA$.  Top right: intersection of flux surfaces with POS as in Figure \ref{fig:fluxsurfaces}, but for a flux rope of greater magnetic helicity.  Sheet-like localization of dipped magnetic fields (brown) represent likely location of prominence formation (see text).} 
\label{fig:nougatvoids}       
\end{figure}

Disk-like or ring-like substructures within cavities are not limited to plasma flows.  Frequent and long-lived manifestations of such substructures are also seen in EUV and SXR emission, sometimes perched atop the prominences like lollypops on their sticks (Figure \ref{fig:nougatvoids}).

\subsubsection{Chewy nougats and central voids}
\label{subsubsec:nougobs}

The first substructure to be noted within cavities were the hot central cores, often referred to as ``chewy nougats'', that sometimes appear in SXR observations \citep{hudson_99} (Figure \ref{fig:nougatvoids} -- top).  Such structures are not transient, but may persist for as long as the cavity is visible at the limb, and indeed  reappear from rotation to rotation for months \citep{hudson_00}.  Nougats may either appear in the form of a disk, or alternatively as a ring nested within the larger-scale cavity, when seen in projection at the limb. By employing the Sun's rotation to extract longitudinal information in a similar manner to that employed in determining the 3D morphology of the larger-scale cavity, the SXR nougat morphology has been modeled as a hot-walled, hollow tube lying within the cavity \citep{reeves_12}.

There is not, generally, a corresponding signature to the SXR nougat at EUV wavelengths; that is, usually only the larger-scale cavity surrounding the nougat is apparent.  However, a similar feature does sometimes occur, in the form of a central low-emission void lying above the prominence with a U-shaped lower boundary (Figure \ref{fig:nougatvoids} -- bottom).  Unlike the nougats, these structures seem to primarily be associated with cavities that are soon to erupt (see Section \ref{subsubsec:teardrop}).

\runinhead{MHD interpretation of substructure: thermodynamic and/or topological interfaces.}  
As discussed in Section \ref{subsubsec:dens}, the thermodynamic properties within a flux rope may vary significantly, depending upon field line length.  An additional factor is field line curvature, which affects how dipped (or flat) a field line is.  This may affect the ability of a prominence to condense and the nature of its dynamics \citep{karpen14_book}.

\begin{figure}[t!]
\center{\includegraphics[scale=.27]{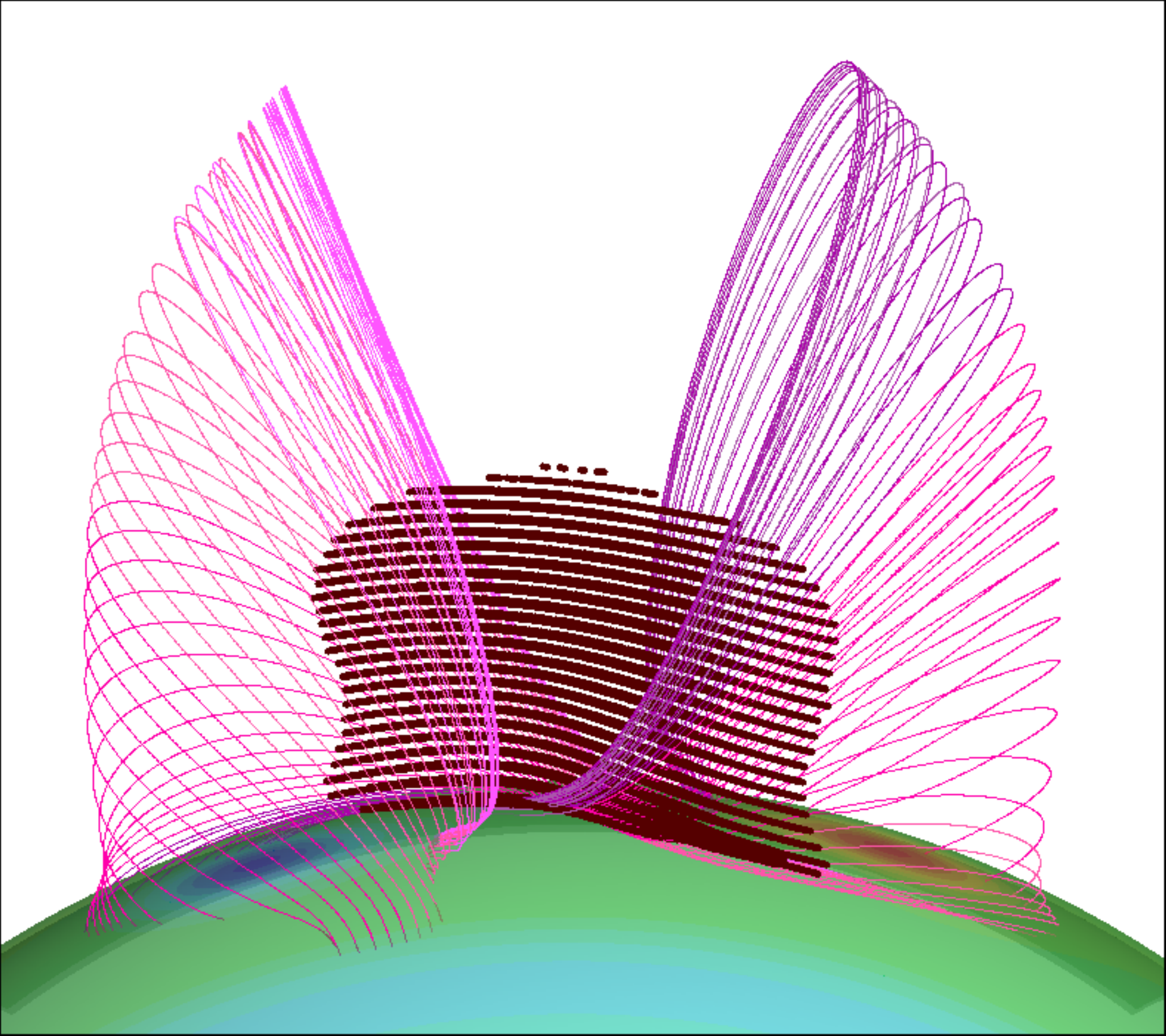}
\includegraphics[scale=.27]{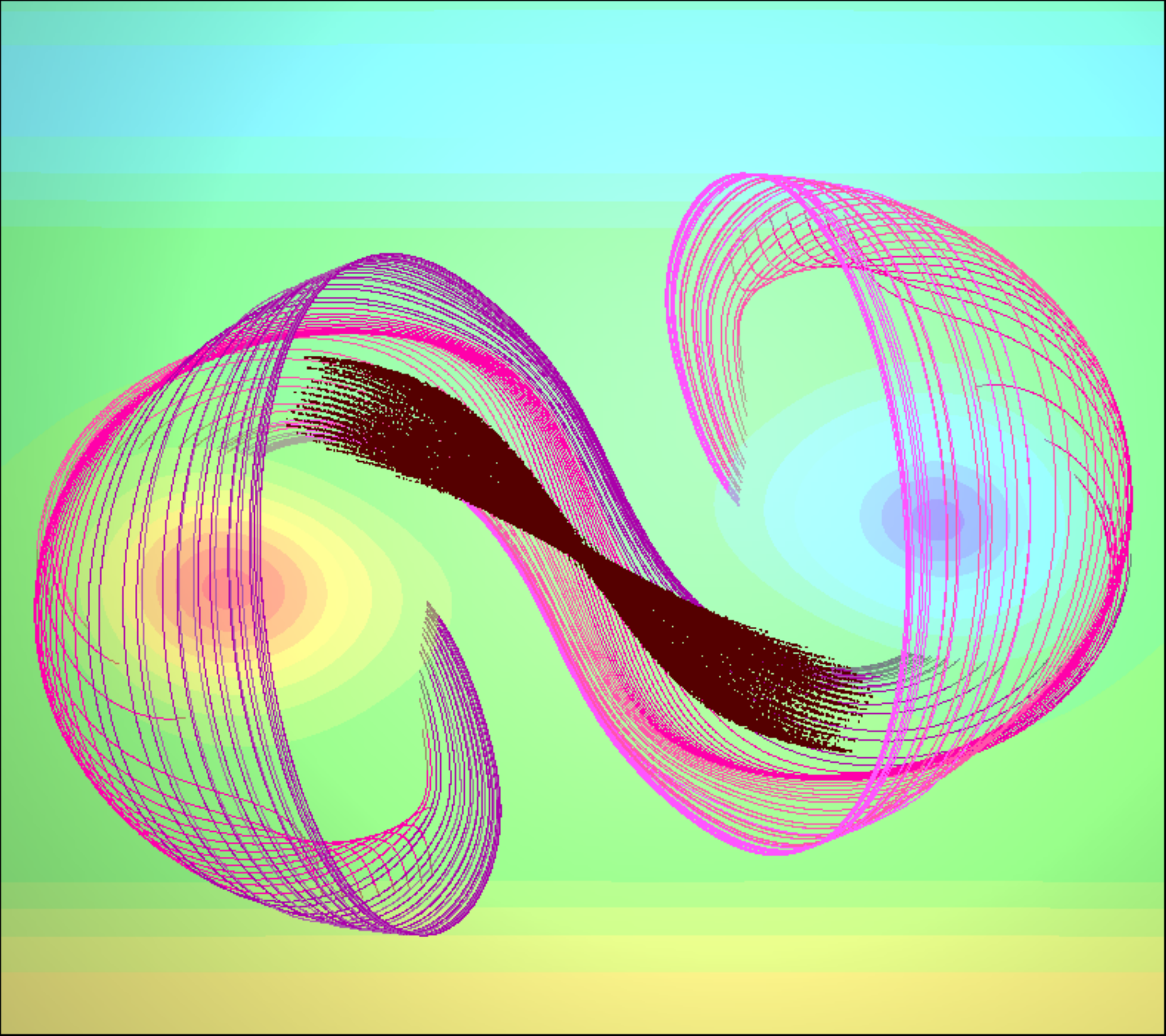}}
\caption{Locus of dips in flux rope is a sheet-like structure, lying above and with fields parallel to underlying neutral line.  The subset of non-arched field lines within the rope extend inward from the outer boundary of the rope (set of purple/pink field lines).  
Brown dots trace out points where the field is dipped relative to the solar radial coordinate, up to a prominence scale height (adapted from \citet{gibfan_06b}).} 
\label{fig:prominence}       
\end{figure}

Consider a magnetic flux rope with field lines that wrap between one and two times about the axis.  
If this rope has an arched axis, only the outermost field lines will possess a dip, while the inner, axial field lines will possess an arched geometry which will be unsuited to supporting prominence mass against gravity. This was demonstrated in Figure \ref{fig:length}, and we now consider how the location of dipped field relates to the flux rope as a whole.   If we fill the dipped field lines to a prominence scale height \citep{Aulanier05a}, we obtain a sheet-like surface made up of multiple threads, lying above the neutral line (Figure \ref{fig:prominence}).  The magnetic fields associated with  these threads run essentially parallel to the underlying neutral line. This is consistent with filament and filament channel observations \citep{martin14_book}.

There are then three distinct thermodynamic regimes likely within a flux rope:

\begin{enumerate}
\item The dipped field lines which lie at the outer boundary of the flux rope, but project into its POS cross section, which might be expected to be denser, dynamic, and multithermal
\item The sheet-like locus of dips of these field lines which would be the most likely place for a quiescent prominence to form (although which dips are filled at any given moment may vary, resulting in an inhomogeneous structure)
\item The central, nondipped (axial) portion of the flux rope, which, due to arched geometry and short fieldline length, would be most likely to manifest the low density of a ``true'' cavity
\end{enumerate}

\begin{figure}[b!]
\center{
\includegraphics[scale=.21]{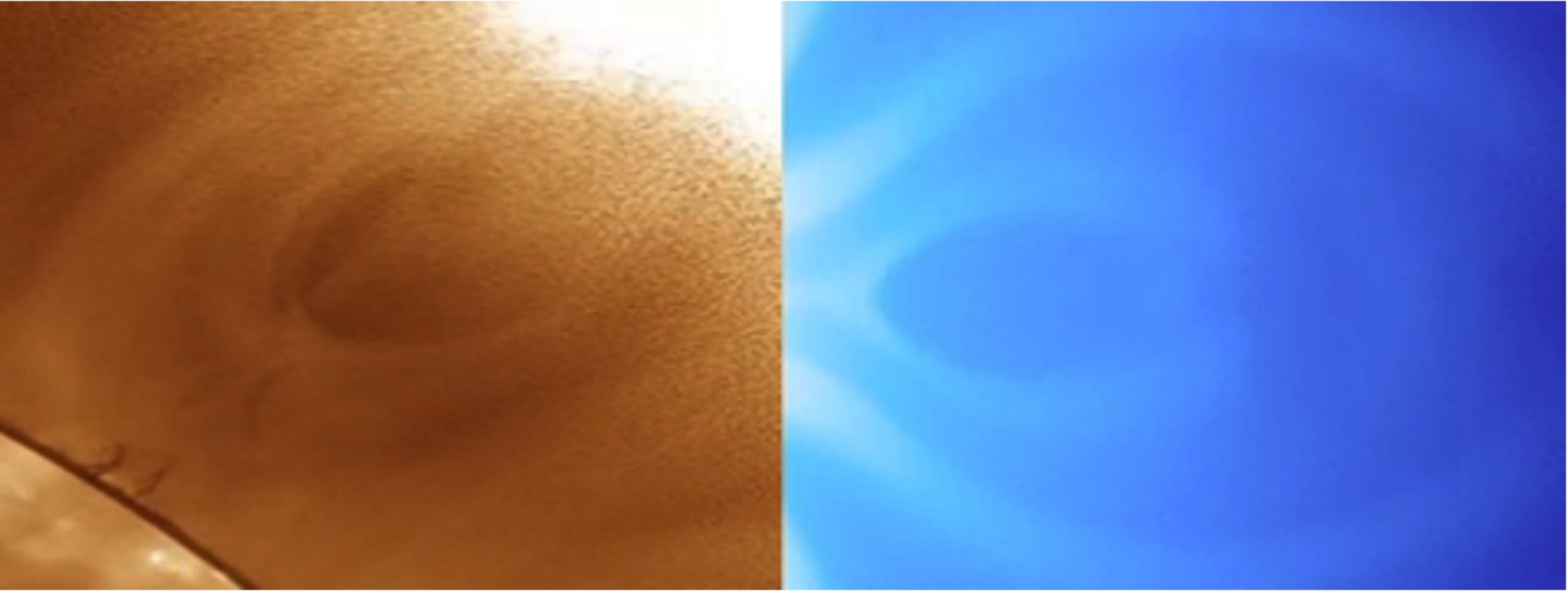}
\includegraphics[scale=.25]{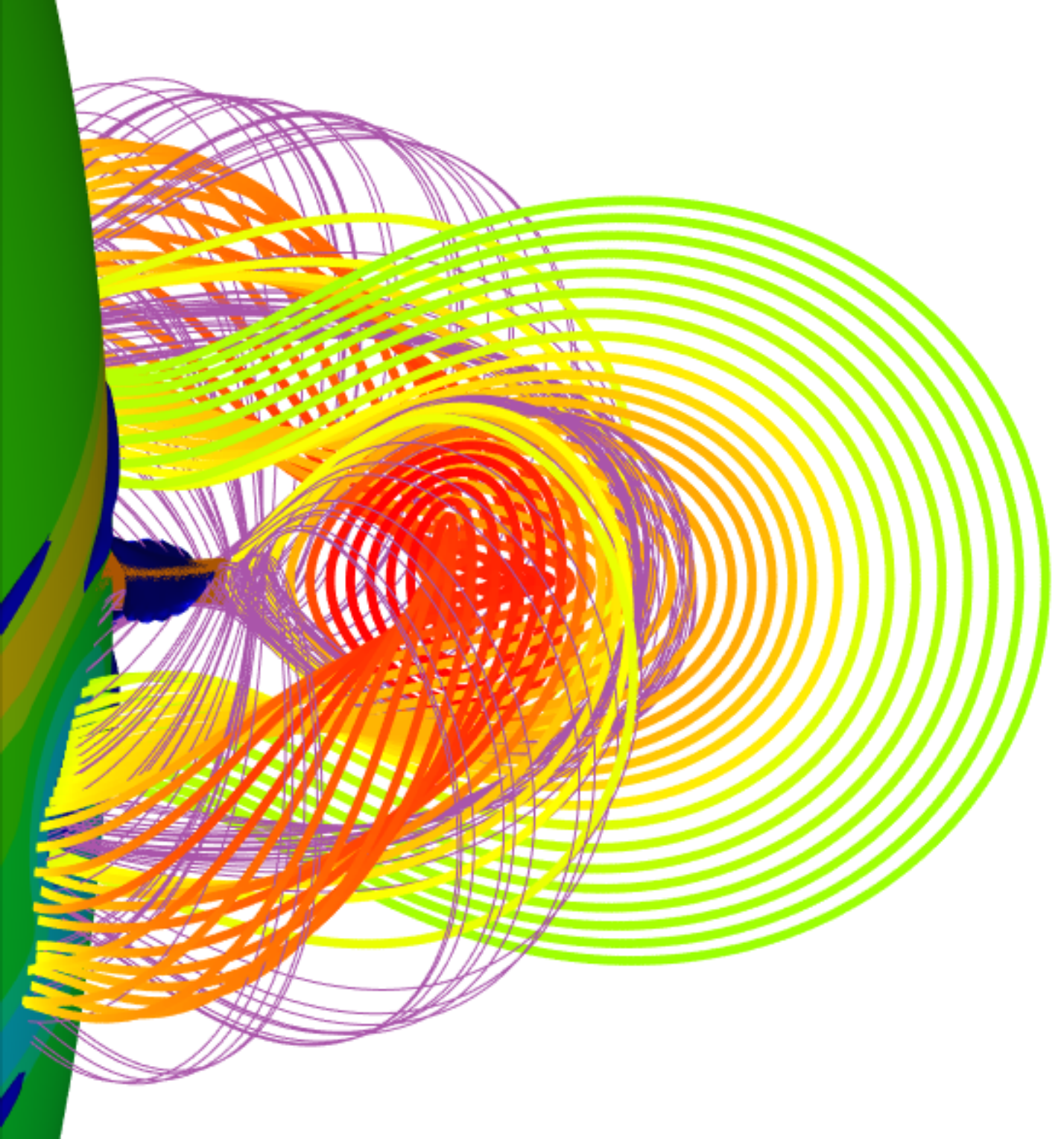}}
\caption{Current sheet formation at flux rope base and associated reconnection may explain hot, low density core within cavity. Left: central void surrounded by U-shaped horns observed atop prominence in hours prior to an eruption (case analyzed by \citet{regnier_11}).  Middle: current sheets surrounding and extending vertically below central void within a simulated flux rope at a late (but still quasistatic) stage of its evolution (from \citet{fan_12}; reproduced by permission of the AAS).  Right: reconnections at the top of this current sheet (dark blue central structure extending up from photosphere, essentially co-localized with prominence dips) lead to heating and flows along central part of rope (temperature indicated by field line color, green-cold, red-hot).} 
\label{fig:ropeeverupt}       
\end{figure}

Current sheets may also form between topologically-distinct regions associated with flux ropes, for example at separatrix surfaces.  One such surface (pink-purple lines of Figure \ref{fig:prominence}) is defined by field lines intersecting the ``bald patch'' of concave-up field at the photospheric polarity inversion line, and may give rise to sigmoid-shaped reconnecting field lines at its outer boundary  \citep{titdem,gibsigobs}.  We can generalize to boundaries where strong gradients in field line length have functionally similar consequences as true magnetic discontinuities -- so-called quasi-separatrix layers, or QSLs \citep{Demoulin96b,fan14_book}.  A QSL can exist at the center of a flux rope, as the upward stretching of the flux rope in response to ongoing helicity/flux input leads to a Hyperbolic Flux Tube (HFT) topology \citep{Titov07}.  Reconnections occuring at the top of this HFT result in a central bundle of recently-reconnected low-density, high-temperature field lines lying above the dipped field lines of a prominence (Figure \ref{fig:ropeeverupt}). This has been proposed as an explanation for the chewy nougat and low density central structures within cavities, at least for pre-eruption observations  \citep{fan_12}.  We will discuss this further in Section \ref{subsubsec:teardrop}.

Lollypop-like structures within cavities thus may arise as a  natural consequence of these thermodynamic and topological divisions.  The prominence, being associated with the sheet of dipped field, would lie below the central, arched field of the rope.  This central region of the field would be thermodynamically, and potentially topologically, separated from the rest of the rope.  It would thus be prone to reconnection at its outer boundary, leading to the hot ring or disk-like structure of the nougat above the prominence.  When the rope axis is at its most arched, perhaps in the days/hours leading up to eruption (Section \ref{subsec:precursor}), this central portion will stretch above the prominence and may appear as a clear, central void.

\begin{svgraybox}
\runinhead{Open Questions}
\begin{itemize}
\item Nougats are known to exist quiescently; is the same true for dark central voids surrounded by U-shaped horns, or do these only occur in the hours leading up to an eruption?
\item Does the formation of an HFT topology inevitably lead to an eruption within a day or two? 
\end{itemize}
\end{svgraybox}

\begin{figure}[t!]
\center{\includegraphics[scale=.6]{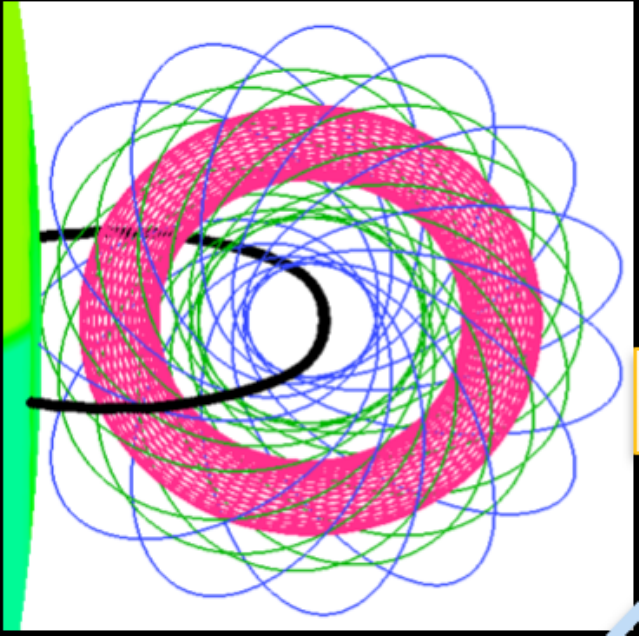}
\includegraphics[scale=2.5]{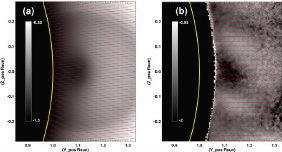}}
\includegraphics[scale=.32]{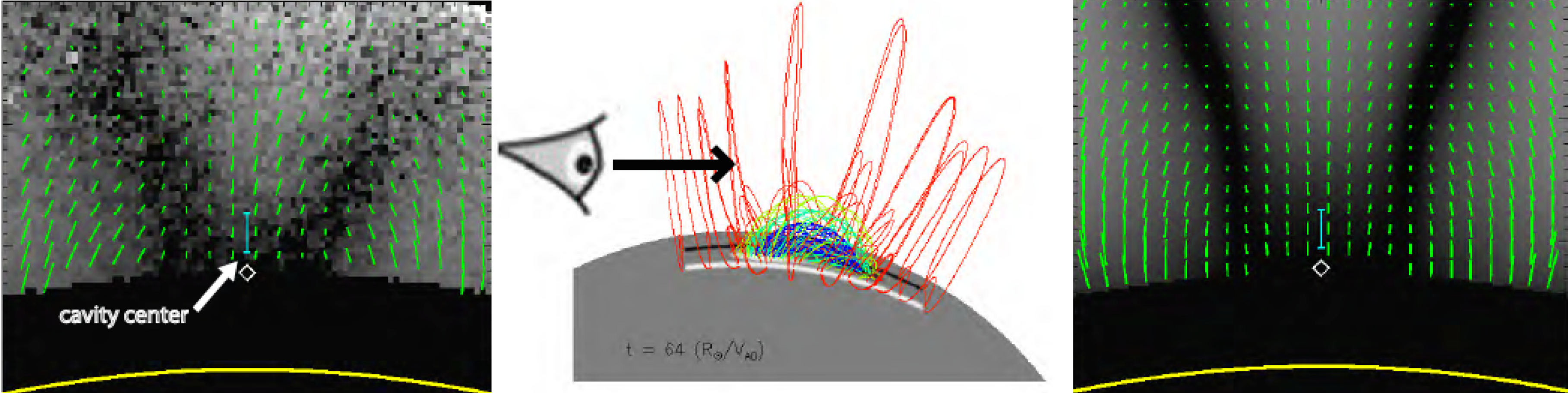}
\caption{Linear polarization measurements of cavities imply non-potential fields.  Top: toroidal (spheromak-like) flux-rope magnetic model (left), forward-modeled to produce synthetic $L/I$ (middle) with bright-ring-like structure similar to that observed within a (non-PCF) cavity by the CoMP telescope (right) (adapted from \citet{dove_11}).  Red linear polarization vectors are essentially radial for both model and data.  Bottom: CoMP $L/I$ observation of lagomorph (rabbit-head) shaped feature within a PCF cavity (left).   Applying a flux-rope model (middle), the LOS-integration results in a clear $L/I$ lagomorph (right). Linear polarization vectors (shown here in green) for model and data show similar deflection from radial at the boundary of the lagomorph (adapted from \citet{ula_13}). } 
\label{fig:lagomorph}       
\end{figure}

\subsection{Polarimetry}
\label{subsec:polobs}

We have focused so far on coronal plasma observations to gain information about cavity magnetic structure.  This approach is a good first step, and justifed in a magnetically-dominated regime where the plasma acts as a tracer of magnetic structure.  However, only part of the field is generally traced out, and the approach is limited by necessary assumptions about the nature of the sub-regions of field that are so traced (are they field lines? dipped-field? current sheets? QSLs?) and the physical mechanisms associated with their visibility (heating? cooling? flows?).  
Therefore, it is important to directly measure magnetic fields within the cavity.  Such measurements are possible using Stokes polarimetry of forbidden infrared (IR) lines where the linear polarization is dominated by the Hanle effect in the saturated regime (see \citet{rachmeler_13}; also \citet{lopez14_book} for a discussion of similar techniques used in the measurement of magnetic fields of prominences).  

The fraction of linearly-polarized light ($L/I=\sqrt{Q^2+U^2}/I$, where $Q$ and $U$ are Stokes vectors) has turned out to be a particularly useful diagnostic for coronal cavities.  It provides a measure of the direction of the magnetic field: linearly-polarized light is strongest where $B$ lies in the plane of the sky, and goes to zero when $B$ lies along the line of sight.  In addition, $L/I$ vanishes when the magnetic field orientation lies at the Van Vleck angle of $54.7^\circ$ to the solar radial.  
As the linear polarization vector crosses this critical angle, its orientation undergoes a $90^\circ$ rotation introducing an ambiguity in addition to the more standard $180^\circ$ one.  The result is a tendency for linear polarization vectors to manifest with a radial orientation (Figure \ref{fig:lagomorph}).  However, the location of nulls in $L/I$ can be used to indicate LOS-directed field and/or Van-Vleck crossings.

The first cavity studied using IR polarimetry was from 2005, and was a large, but not PCF cavity.  The linear polarization associated with the cavity was in the form of a bright ring, with a dark core and surrounding dark ring (Figure \ref{fig:lagomorph} top).
Over the past few years, synoptic observations have been available and have indicated that a much more common linear polarization signal associated with cavities, in particular PCF cavities, is a structure akin to that of a rabbit's head (``lagomorph'') (Figure \ref{fig:lagomorph} bottom) \citep{ula_13}.  Linear-polarization lagomorphs generally scale with cavity size \citep{ula_14}.

\runinhead{MHD interpretation of lagomorphs: axial field surrounded by poloidal field.}
Because of the complexities intrinsic to spectropolarimetry in combination with line-of-sight integration effects, forward modeling represents a useful means of interpreting  observations.  Using a quantum-electrodynamical formulation to synthesize  coronal Stokes parameters \citep{casinijudge_99}, the general sensitivity of $L/I$ to the presence of coronal currents has been demonstrated \citep{judge_06}. Moreover, forward-modeled $L/I$ for a flux rope with a toroidal (spheromak-like) magnetic topology \citep{giblow_98} was shown to result in the type of nested rings observed in the 2005 cavity \citep{dove_11} (Figure \ref{fig:lagomorph}).

We therefore are led to consider what $L/I$ structure would result from a simpler, cylindrical flux rope.  Figure \ref{fig:lagomorph} (bottom) illustrates that, even considering LOS-integration effects, a lagomorph is precisely what we should expect to observe.  A simple arcade of magnetic field lines, perpendicular to and viewed along the neutral line, will result in a V-shape because of the Van Vleck inversion where the magnetic field is oriented $54.7^\circ$ to the local vertical (solar radial) direction.  A flux rope oriented along the LOS will possess an axial component that will distort the location of the Van Vleck nulls, and lead to vanishing $L/I$ at its axis.  LOS-integration of a curved flux rope will smear out these nulls to some extent, but if the rope is sufficiently long and straight (which is likely if it manifests as a distinct cavity), the result will be a lagomorph \citep{ula_13,rachmeler_13}.

The magnetic flux rope model is thus consistent with the observed $L/I$ lagomorphs, but is it unique in this consistency?  In particular, how would a sheared-arcade configuration, in which field lines exhibit writhe but do not wrap around an axis, appear?  Forward-modeling \citep{rachmeler_13} shows that LOS-integrated $L/I$ in the upper portions of the flux-rope and sheared-arcade models are very similar, assuming both models are surrounded by a simple arcade that produces the V-shaped rabbit's "ears".  Since both models possess an axial component of field beneath this arcade, they also predict more parallel Van-Vleck inversions outlining the rabbit's "head".  The flux-rope model however generally predicts a central darkening, since the axial field peaks at rope center. On the other hand,  the sheared-arcade model has axial field which extends down to the photosphere. Therefore, disambiguation should be possible using observations of the polarization vector beneath the cavity center, or of the radial profile of circular polarization (Stokes $V$), which is proportionate to the LOS magnetic field \citep{rachmeler_13}.    However, telescopes with lower occulting disks and larger light-gathering capacity than currently available may be required for such a study.

\begin{svgraybox}
\runinhead{Open Questions}
\begin{itemize}
\item How does circular polarization (Stokes $V$) vary within the cavity?
\item What is the orientation of linear polarization below cavity center?
\end{itemize}
\end{svgraybox}

\begin{figure}[t!]
\includegraphics[scale=.5]{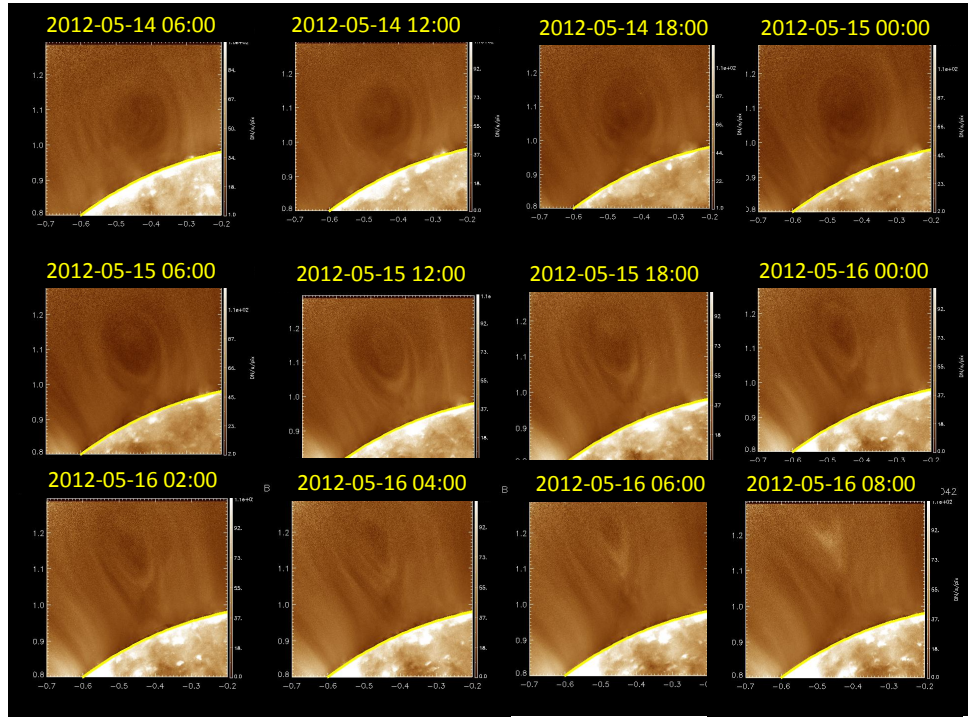}
\caption{A long-lived cavity that eventually erupts, and exhibits a slow ``activation'' phase in the hours leading up to eruption. SDO/AIA $193~\AA$ observations.} 
\label{fig:bodily}       
\end{figure}

\section{Erupting cavities and their precursors and predictors}
\label{sec:erupt}

Coronal mass ejections (CMEs) often exhibit a three-part structure of leading edge, following cavity, and entrained prominence core \citep{illhund,webb14_book}.  Such CMEs are generally pointed to as evidence for "flux rope CMEs" \citep{giblow_98,dere,zhanglow_05}.  Controversy remains as to whether the flux rope structure existed prior to eruption, or was formed during eruption \citep{fan14_book,gopal14_book}.   In many cases, the CME source region is low-lying, or surrounded by complex structures, making it difficult to analyze the pre-eruption configuration using observations as described in Section \ref{sec:quiescent}. Cases where a clearly pre-existing cavity erupts (Figure \ref{fig:bodily}) are useful for studying this question in a straightforward fashion, at least as it pertains to a subset of CMEs.

\subsection{Erupting cavities}
\label{subsec:bodily}

Figure \ref{fig:bodily} shows a cavity which was visible for at least 48 hours before eruption and essentially unchanging for at least 24 hours before signs of activation.  It is an example of what we refer to as a ``bodily-erupting" cavity, in which the precursor cavity morphology can be tracked throughout the eruption.  Indeed, bodily-erupting cavities have been tracked from pre-eruption through the corona and out into the heliosphere \citep{deforest_13}.

In the survey of $> 100$ EUV cavities \citep{forland_13} described in Section \ref{sec:quiescent}, eruptions were witnessed in approximately a third of the cavities monitored, and bodily-erupting cavities were evident for about a quarter of all cavities.   The white-light survey of cavities produced similar results \citep{gibcav} -- roughly a third were observed to erupt within 48 hours (24 out of 74 cases).  Establishing which of these white-light cavities erupted bodily required monitoring the eruption onset.  Due to the limited observing time window -- a duty cycle that sampled about $1/4$ of the hours in a given day, and was further reduced by weather -- only four could be so monitored, and all of these were observed to bodily erupt. Considering the finite length of cavities and the fact that they can only be observed at the solar limbs, a rough extrapolation implies that all cavities eventually erupt, and the majority of these bodily erupt. 

Further insight can be gained from a study \citep{maricic_09} which considered 18 3-part CMEs in which a leading edge (LE) and erupting prominence (EP) were visible from pre-eruption onwards.  Note that for many of the cases (which included active region related eruptions) no pre-existing quiescent cavity could be distinguished: in such cases the appearance of a bright loop during the hours before an eruption was sufficient to establish the LE.   The acceleration properties of the eruptions were analyzed, and bodily eruption quantified by tracking the kinematics of the LE vs. the EP.  In the majority of the cases studied, the LE and EP acceleration began almost simultaneously (within 20 min).  However, in two events the EP acceleration began considerably earlier than the LE acceleration ($>40$ min) and in two events the LE acceleration started $>50$ min before the EP acceleration. 

Another study tracked the eruption of a pre-existing cavity, and found that although it clearly erupted, its associated prominence did not appear to escape with it  \citep{liu_07}.  This was interpreted as a partial eruption, akin to the phenomenon where prominences seem to split in two during a CME -- one part escaping, one part falling back to the Sun \citep{gibfan_06a}.

\runinhead{MHD interpretation of erupting cavities: destabilization and loss of equilibrium.}
As discussed in \citet{fan14_book}, CMEs may be triggered by an ideal process (e.g., the kink and torus instabilities), or by a reconnection-driven process (e.g., the breakout model).  In the case of a reconnection-driven trigger, no pre-existing flux rope is required as it may form as a consequence of the eruption (although a breakout-type reconnection can also operate above a pre-existing flux rope \citep{torok_11}).  For cases in which the prominence and leading edge eruption are not synchronized, thus, no bodily eruption, it seems plausible that the rope indeed forms or at least is greatly transformed during eruption.  However, for the majority of cases, a cavity or LE and following prominence exist prior to the eruption and are synchronized in their motion from its beginning.  These observations provide compelling evidence for pre-existing flux ropes that are destabilized and undergo loss of equilibrium.  

\begin{svgraybox}
\runinhead{Open Questions}
\begin{itemize}
\item To what extent does the behavior exhibited by bodily-erupting cavities extend to other CMEs?
\end{itemize}
\end{svgraybox}

\subsection{Precursors and predictors of eruption}
\label{subsec:precursor}

Figure \ref{fig:bodily} demonstrates some characteristic behavior of cavities in the hours leading up to eruption.  This may be classified as ``cavity activation'', and is likely to be related to similar manifestations in quiescent prominences in the hours prior to their eruption  \citep{martin14_book} as well as pre-eruption behavior of so-called ``streamer blowouts'' \citep{webb14_book}.  In particular, cavities appear to slowly increase in height, to narrow and become more teardrop-shaped, and to exhibit more sharply defined sub-structure prior to eruption. Even before activation, there is evidence that quasistatic evolution results in observational characteristics that may be used as predictors of impending eruption.

\subsubsection{Cavity height and slow rise}
\label{subsubsec:height}

The hours leading up to the impulsive phase of a cavity eruption are often marked by a slow rise of the cavity and its associated prominence.  An example of rise speeds for the two phases of a particular cavity eruption was found to be  $0.6~km~s^{-1}$ for the slow-rise phase and $25~km~s^{-1}$ (average) for the subsequent impulsive phase \citep{regnier_11}.  

Before any measurable slow-rise phase, however, the height of the cavity may provide a clue to its likelihood for eruption.  An absolute upper limit to cavity height appears to be approximately $1.6~\Rsun$ ($0.6~\Rsun$ above the solar surface), beyond which no non-erupting cavities were observed in white light \citep{gibcav}.
Due to selection effects discussed in Section \ref{subsec:ubiqobs}, EUV cavities in general lie below this height (median value $1.2$ solar radii).  Nevertheless, they exhibit a marked tendency for higher cavities to be more likely to erupt: the average cavity center height was $25\%$ higher for cavities that erupted (measured prior to slow rise) than for a baseline set of non-eruptive cavities  \citep{forland_13}.  This is no doubt related to observations indicating an upper limit for prominence height \citep{munro_79,filippov_01,liu_12}.


\subsubsection{Morphology and substructure}
\label{subsubsec:teardrop}

The most obvious sign of cavity activation in Figure \ref{fig:bodily} is the bright substructure that forms by 06:00, initially as U-shaped ``horns'' above the prominence (see Section \ref{subsec:flowsub}), and progressing to a very narrow ellipse within the broader cavity envelope.  The entire cavity likewise narrows during the slow-rise phase of the eruption.  

This behavior is mirrored in the characteristics of pre-activation but soon-to-erupt cavities relative to the baseline, non-eruptive cavities.  In Section \ref{subsec:morphobs}, we discussed the tendency towards prolateness, or narrowness of the elliptical cross sections of cavities.  This tendency is even more pronounced for the subset of cavities that erupt than for the baseline, non-eruptive cavities.  In fact, the pre-activation cavities may be better characterized as possessing a teardrop, rather than elliptical, shape.  Categorizing the EUV cavities by morphology, the likelihood of eruption of teardrop-shaped cavities was 68\%,  as compared to 23\% for more elliptical cavities, and 10\% for cavities that were best described as semicircular \citep{forland_13}.  
Due to the presence of an occulting disk, the full shape of  white-light cavities is not generally measured, but a quality referred to as ``necking'' can be noted when cavities have narrower bases than tops.  In the white-light cavity survey, 10/10 cases of cavities which erupted within 24 hours exhibited necking, vs 25/99 of the entire sample \citep{gibcav}.


\runinhead{MHD interpretation of precursors and predictors: topological changes leading up to ideal instability.}
The absolute upper limit on cavity height, as manifested by white light cavities, is undoubtedly related to the upper limit on the spatial scale of closed field possible in a corona before it ultimately opens into the solar wind, and may represent a limit beyond which cavities are gravitationally unstable.  The smaller EUV cavities, however, lie well below this global limit, so the association of eruptions with increased cavity height may require interpretation in terms of purely magnetic forces.  Indeed, the height of pre-CME cavities taken in conjunction with their narrowness and teardrop shape is consistent with the formation of a current sheet at the base of the rope (e.g., Figure \ref{fig:ropeeverupt}, middle), presenting an intriguing clue as to how magnetic evolution might lead up to eruption.  

The cavity may be thought of potentially having three stages in its evolution.
The first stage is the long-lasting period of time where a cavity exists essentially quiescently, without erupting.  This stage may nevertheless represent a sequence of quasistatic equilibria, in which minimum energy states are continually updated as the lower boundary inputs magnetic helicity through flows or flux emergence over the course of days and weeks.  The transition to the second stage occurs when a current sheet forms beneath the cavity and the topology changes from a flux rope grazing the photosphere (bald-patch) to one with an HFT beneath it (QSL) (see discussion in Section \ref{subsubsec:nougobs}, and in \citet{fan14_book}).  At this point evolution may continue in a quasistatic fashion, but a fuse has been lit and the tether-cutting reconnections at the current sheet increase twist at the core of the flux rope (Figure \ref{fig:ropeeverupt}, right), creating substructure in the cavity and a slow rise of the rope axis.  The third and final stage is dynamic, with eruption possibly triggered by the kink or torus instability brought on by increasing twist and/or rope height.  Alternatively, the slow rise of the rope may push it into topologically distinct, overlying or adjacent magnetic fields, resulting in breakout-type reconnection.  This could also occur in a manner that skips the middle stage of evolution, if, for example, the evolution of surrounding fields forces reconnections and drives a ``sympathetic'' eruption \citep{torok_11}.

\begin{svgraybox}
\runinhead{Open Questions}
\begin{itemize}
\item Can we establish which CMEs are predominantly ideal-instability-driven, and which are reconnection-driven, and whether certain regions (PCFs, active regions) are more likely to be one than the other?
\item As we begin to measure the coronal fields themselves, is there a property: helicity, free energy, complexity, topology -- that we could measure to tell us that eruption is inevitable?
\end{itemize}
\end{svgraybox}


\section{Conclusions}
\label{sec:conclusions}

Cavities hold unique clues to understanding the nature of pre-CME equilibria and the mechanisms that trigger their loss.  They represent the bulk of the combined erupting prominence-cavity volume, so it is their magnetic structure that may map best to the magnetic cloud passing the Earth \citep{gopal14_book,lugaz14_book}.  In this paper we have presented a set of observations that offer strong physical insights into the nature of this magnetic structure.

Cavities are {\it ubiquitous} - at least for the longitudinally-extended filaments (e.g., PCFs) that our observations are necessarily biased towards.   They have a {\it croissant-like morphology},  have {\it low density} (but not as low as a coronal hole), and are {\it multithermal}, at least in projection.  Flows are observed that {\it spatially and temporally link prominence and cavity}.  These flows sometimes take the form of {\it swirling motions} in the plane of the sky, and as {\it nested rings} as measured in LOS velocity.  {\it Disk-like or ring-like sub-structure} is often seen at the center of the cavity and lying above the prominence like a lollypop on a stick.  Finally, recent Stokes polarimetric observations of coronal magnetic field provide direct evidence of field oriented parallel to the underlying neutral line, {\it at heights well above the prominence, and corresponding to the height of the cavity center}.  

Cavities erupt as CMEs, and the majority appear to exhibit {\it bodily eruption}.  Cavities, like filaments, may be activated prior to eruption and show {\it slow rise, narrowing, and enhanced substructure}.  Even before activation, {\it cavity height and teardrop morphology are good predictors of impending eruption}.

Any model of the magnetic structure of a cavity must satisfy all of these observational constraints.  We have argued in this paper that a magnetic flux rope plausibly does so.  In particular it is difficult to see how observations such as the nested bullseye LOS flows could be explained without invoking the toroidal flux surfaces of a flux rope.  However, we have also highlighted open questions throughout our review, chief among them concerning 
the degree to which what we have learned about PCF cavities extends to all prominences and CME source regions. 
Mysteries remain, but cavities continue to yield intriguing glimpses into the hearts of CMEs, from pre-event out into the heliosphere.

\begin{acknowledgement}
The National Center for Atmospheric Research is sponsored by the National Science Foundation.  AIA data courtesy of NASA/SDO and the AIA, EVE, and HMI science teams.  $H\alpha$ image courtesy of Big Bear Solar Observatory/New Jersey Insitute of Technology.  Hinode is a Japanese mission developed and launched by ISAS/JAXA, with NAOJ as domestic partner and NASA and STFC (UK) as international partners. It is operated by these agencies in co-operation with ESA and NSC (Norway).
Much of the work presented here directly relates to, or benefited greatly from research undertaken by the International Space Science Institute (ISSI) international teams on coronal cavities (2008-2010) and coronal magnetism (2013-2014). I am indebted to all of the members of both of these teams, particularly Urszula Bak-Steslicka, Terry Kucera, Laurel Rachmeler, Kathy Reeves, and Don Schmit.  In addition, I thank Tom Berger, Giuliana de Toma, Yuhong Fan, Blake Forland, Jim Fuller, Judy Karpen, Jim Klimchuk, Olga Panasenco, Marco Velli, and especially B. C. Low for many helpful discussions.
 \end{acknowledgement}


\begin{thebibliography}{96}
\providecommand{\natexlab}[1]{#1}
\providecommand{\url}[1]{{#1}}
\providecommand{\urlprefix}{URL }
\expandafter\ifx\csname urlstyle\endcsname\relax
  \providecommand{\doi}[1]{DOI~\discretionary{}{}{}#1}\else
  \providecommand{\doi}{DOI~\discretionary{}{}{}\begingroup
  \urlstyle{rm}\Url}\fi
\providecommand{\eprint}[2][]{\url{#2}}

\bibitem[{Aulanier et~al(2005)Aulanier, D{\'e}moulin, and
  Grappin}]{Aulanier05a}
Aulanier G, D{\'e}moulin P, Grappin R (2005) {Equilibrium and observational
  properties of line-tied twisted flux tubes}. Astronomy and Astrophysics
  430:1067

\bibitem[{Ballester(2014)}]{ballester14_book}
Ballester JL (2014) Magnetism and dynamics of prominences: Mhd waves. In:
  Engvold, Vial (eds) Solar Prominences, Springer, p~00

\bibitem[{{B{\c a}k-St{\c e}{\'s}licka} et~al(2013){B{\c a}k-St{\c
  e}{\'s}licka}, Gibson, Fan, Bethge, Forland, and Rachmeler}]{ula_13}
{B{\c a}k-St{\c e}{\'s}licka} U, Gibson SE, Fan Y, Bethge C, Forland B,
  Rachmeler LA (2013) Twisted magnetic structure of solar prominence cavities:
  New observational signature revealed by coronal magnetometry. {\it Astrophys
  J} in press; Arxiv 13047388 770:28

\bibitem[{{B{\c a}k-St{\c e}{\'s}licka} et~al(2014){B{\c a}k-St{\c
  e}{\'s}licka}, Gibson, Fan, Bethge, Forland, and Rachmeler}]{ula_14}
{B{\c a}k-St{\c e}{\'s}licka} U, Gibson SE, Fan Y, Bethge C, Forland B,
  Rachmeler LA (2014) The spatial relation between euv cavities and linear
  polarization signatures. IAU S300 Proceedings

\bibitem[{{Berger}(1984)}]{Berger84}
{Berger} MA (1984) {Rigorous new limits on magnetic helicity dissipation in the
  solar corona}. Geophysical and Astrophysical Fluid Dynamics 30:79--104,
  \doi{10.1080/03091928408210078}

\bibitem[{{Berger}(2012)}]{bergerasp_2012}
{Berger} T (2012) {Quiescent Prominence Dynamics: An Update on Hinode/SOT
  Discoveries}. In: {Sekii} T, {Watanabe} T, {Sakurai} T (eds) Hinode-3: The
  3rd Hinode Science Meeting, Astronomical Society of the Pacific Conference
  Series, vol 454, p~79

\bibitem[{{Berger} et~al(2011){Berger}, {Testa}, {Hillier}, {Boerner}, {Low},
  {Shibata}, {Schrijver}, {Tarbell}, and
  {Title}}]{Berger.hot-bubble.2011Natur.472..197B}
{Berger} T, {Testa} P, {Hillier} A, {Boerner} P, {Low} BC, {Shibata} K,
  {Schrijver} C, {Tarbell} T, {Title} A (2011) {Magneto-thermal convection in
  solar prominences}. Nature 472:197--200, \doi{10.1038/nature09925}

\bibitem[{{Berger} et~al(2008){Berger}, {Shine}, {Slater}, {Tarbell}, {Title},
  {Okamoto}, {Ichimoto}, {Katsukawa}, {Suematsu}, {Tsuneta}, {Lites}, and
  {Shimizu}}]{BergerT.promin-plume.2008ApJ...676L..89B}
{Berger} TE, {Shine} RA, {Slater} GL, {Tarbell} TD, {Title} AM, {Okamoto} TJ,
  {Ichimoto} K, {Katsukawa} Y, {Suematsu} Y, {Tsuneta} S, {Lites} BW, {Shimizu}
  T (2008) {Hinode SOT Observations of Solar Quiescent Prominence Dynamics}.
  Astrophys J Lett 676:L89--L92, \doi{10.1086/587171}

\bibitem[{{Berger} et~al(2010){Berger}, {Slater}, {Hurlburt}, {Shine},
  {Tarbell}, {Title}, {Lites}, {Okamoto}, {Ichimoto}, {Katsukawa}, {Magara},
  {Suematsu}, and {Shimizu}}]{BergerT.bubble-RT-instable.2010ApJ...716.1288B}
{Berger} TE, {Slater} G, {Hurlburt} N, {Shine} R, {Tarbell} T, {Title} A,
  {Lites} BW, {Okamoto} TJ, {Ichimoto} K, {Katsukawa} Y, {Magara} T, {Suematsu}
  Y, {Shimizu} T (2010) {Quiescent Prominence Dynamics Observed with the Hinode
  Solar Optical Telescope. I. Turbulent Upflow Plumes}. Astrophys J
  716:1288--1307, \doi{10.1088/0004-637X/716/2/1288}

\bibitem[{{Berger} et~al(2012){Berger}, {Liu}, and
  {Low}}]{Berger.Liu.cavity.condense.2012ApJ...758L..37B}
{Berger} TE, {Liu} W, {Low} BC (2012) {SDO/AIA Detection of Solar Prominence
  Formation within a Coronal Cavity}. Astrophys J Lett 758:L37,
  \doi{10.1088/2041-8205/758/2/L37}, \eprint{1208.3431}

\bibitem[{{Casini} and {Judge}(1999)}]{casinijudge_99}
{Casini} R, {Judge} PG (1999) {Spectral Lines for Polarization Measurements of
  the Coronal Magnetic Field. II. Consistent Treatment of the Stokes Vector
  forMagnetic-Dipole Transitions}. AMP 522:524, \doi{10.1086/307629}

\bibitem[{{De Pontieu} et~al(2011){De Pontieu}, {McIntosh}, {Carlsson},
  {Hansteen}, {Tarbell}, {Boerner}, {Martinez-Sykora}, {Schrijver}, and
  {Title}}]{depontieu_11}
{De Pontieu} B, {McIntosh} SW, {Carlsson} M, {Hansteen} VH, {Tarbell} TD,
  {Boerner} P, {Martinez-Sykora} J, {Schrijver} CJ, {Title} AM (2011) {The
  Origins of Hot Plasma in the Solar Corona}. Science 331:55--,
  \doi{10.1126/science.1197738}

\bibitem[{{de Toma} et~al(2008){de Toma}, {Casini}, {Burkepile}, and
  {Low}}]{detoma_08}
{de Toma} G, {Casini} R, {Burkepile} JT, {Low} BC (2008) {Rise of a Dark Bubble
  through a Quiescent Prominence}. Astrophys J Lett 687:L123--L126,
  \doi{10.1086/593326}

\bibitem[{DeForest et~al(2013)DeForest, Howard, and McComas}]{deforest_13}
DeForest C, Howard TA, McComas DJ (2013) Tracking coronal features from the low
  corona to earth: A quantitative analysis of the 2008-dec-12 cme. ApJ,
  submitted

\bibitem[{D{\'e}moulin et~al(1996)D{\'e}moulin, Priest, and
  Lonie}]{Demoulin96b}
D{\'e}moulin P, Priest ER, Lonie DP (1996) {Three-dimensional magnetic
  reconnection without null points 2. Application to twisted flux tubes}.
  Journal of Geophysical Research 101:7631

\bibitem[{Dere et~al(1999)Dere, Brueckner, Howard, Michels, and
  Delaboudiniere}]{dere}
Dere KP, Brueckner GE, Howard RA, Michels DJ, Delaboudiniere JP (1999) Lasco
  and eit observations of helical structure in coronal mass ejections.
  Astrophys J 492:804

\bibitem[{Dove et~al(2011)Dove, Gibson, Rachmeler, Tomczyk, and
  Judge}]{dove_11}
Dove J, Gibson S, Rachmeler LA, Tomczyk S, Judge P (2011) A ring of polarized
  light: Evidence for twisted coronal magnetism in cavities. Astrophys J 731:1

\bibitem[{{Dud{\'{\i}}k} et~al(2012){Dud{\'{\i}}k}, {Aulanier}, {Schmieder},
  {Zapi{\'o}r}, and {Heinzel}}]{dudik_12}
{Dud{\'{\i}}k} J, {Aulanier} G, {Schmieder} B, {Zapi{\'o}r} M, {Heinzel} P
  (2012) {Magnetic Topology of Bubbles in Quiescent Prominences}. Astrophys J
  761:9, \doi{10.1088/0004-637X/761/1/9}

\bibitem[{Engvold(1989)}]{eng89}
Engvold O (1989) In: Priest ER (ed) Dynamics and Structures of Quiescent
  Prominences, D. Reidel Publ Co., p~47

\bibitem[{Engvold(2014)}]{engvold14_book}
Engvold O (2014) In: Engvold, Vial (eds) Solar Prominences, Springer, p~00

\bibitem[{{Fan}(2012)}]{fan_12}
{Fan} Y (2012) {Thermal Signatures of Tether-cutting Reconnections in
  Pre-eruption Coronal Flux Ropes: Hot Central Voids in Coronal Cavities}.
  Astrophys J 758:60, \doi{10.1088/0004-637X/758/1/60}, \eprint{1205.1028}

\bibitem[{Fan(2014)}]{fan14_book}
Fan Y (2014) Magnetism and dynamics of prominences: Mhd equilibria and triggers
  for eruption. In: Engvold, Vial (eds) Solar Prominences, Springer, p~00

\bibitem[{{Filippov} and {Den}(2001)}]{filippov_01}
{Filippov} BP, {Den} OG (2001) {A critical height of quiescent prominences
  before eruption}. J Geophys Res 106:25,177--25,184,
  \doi{10.1029/2000JA004002}

\bibitem[{Forland et~al(2013)Forland, Gibson, Dove, Rachmeler, and
  Fan}]{forland_13}
Forland BF, Gibson SE, Dove JB, Rachmeler LA, Fan Y (2013) Coronal cavity
  survey: morphological clues to eruptive magnetic topologies. Solar Phys, in
  press

\bibitem[{Fuller and Gibson(2009)}]{fuller_09}
Fuller J, Gibson SE (2009) A survey of coronal cavity density profiles.
  Astrophys J 700:1205

\bibitem[{Fuller et~al(2008)Fuller, Gibson, de~Toma, and Fan}]{fuller_08}
Fuller J, Gibson SE, de~Toma G, Fan Y (2008) Observing the unobservable?
  modeling coronal cavity density. Astrophys J 678:515

\bibitem[{Gibson(2014)}]{gibsoniau_14}
Gibson SE (2014) Magnetism and the invisible man: The mysteries of coronal
  cavities. IAU S300 Proceedings

\bibitem[{Gibson and Fan(2006{\natexlab{a}})}]{gibfan_06b}
Gibson SE, Fan Y (2006{\natexlab{a}}) Coronal prominence structure and
  dynamics: A magnetic flux rope interpretation. J Geophys Res 111,
  \doi{10.1029/2006JA011871}

\bibitem[{Gibson and Fan(2006{\natexlab{b}})}]{gibfan_06a}
Gibson SE, Fan Y (2006{\natexlab{b}}) The partial expulsion of a magnetic flux
  rope. Astrophys J Lett 637:65

\bibitem[{Gibson and Low(1998)}]{giblow_98}
Gibson SE, Low BC (1998) A time-dependent three-dimensional magnetohydrodynamic
  model of the coronal mass ejection. Astrophys J 493:460

\bibitem[{Gibson et~al(2002)Gibson, Fletcher, Del~Zanna, Pike, Mason, Mandrini,
  Demoulin, Gilbert, Burkepile, Holzer, Alexander, Liu, Nitta, Qiu, Schmieder,
  and Thompson}]{gibsigobs}
Gibson SE, Fletcher L, Del~Zanna G, Pike CD, Mason HE, Mandrini CH, Demoulin P,
  Gilbert H, Burkepile J, Holzer T, Alexander D, Liu Y, Nitta N, Qiu J,
  Schmieder B, Thompson BJ (2002) The structure and evolution of a sigmoidal
  active region. Astrophys J 574:265

\bibitem[{Gibson et~al(2006)Gibson, Foster, Burkepile, de~Toma, and
  Stanger}]{gibcav}
Gibson SE, Foster D, Burkepile J, de~Toma G, Stanger A (2006) The calm before
  the storm: the link between quiescent cavities and cmes. Astrophys J 641:590

\bibitem[{Gibson et~al(2010)Gibson, Kucera, Rastawicki, Dove, de~Toma, Hao,
  Hill, Hudson, Marque, McIntosh, Rachmeler, Reeves, Schmieder, Schmit, Seaton,
  Sterling, Tripathi, Williams, and Zhang}]{gibson_10}
Gibson SE, Kucera TA, Rastawicki D, Dove J, de~Toma G, Hao J, Hill S, Hudson
  HS, Marque C, McIntosh PS, Rachmeler L, Reeves KK, Schmieder B, Schmit DJ,
  Seaton DB, Sterling AC, Tripathi D, Williams DR, Zhang M (2010)
  Three-dimensional morphology of a coronal prominence cavity. Astrophys J
  723:1133

\bibitem[{Gopalswamy(2014)}]{gopal14_book}
Gopalswamy N (2014) Eruptive prominences and their impact on the earth: The
  dynamic phenomenon. In: Engvold, Vial (eds) Solar Prominences, Springer, p~00

\bibitem[{Guhathakurta et~al(1992)Guhathakurta, Rottman, Fisher, Orrall, and
  Altrock}]{guhaetal_92}
Guhathakurta M, Rottman GJ, Fisher RR, Orrall FQ, Altrock RC (1992) Coronal
  density and temperature structure from coordinated observations associated
  with the total solar eclipse of 1988 march 18. Astrophys J 388:633

\bibitem[{Habbal et~al(2010)Habbal, Druckmueller, Morgan, Scholl, Rusin, Daw,
  Johnson, and Arndt}]{habbalcav}
Habbal SR, Druckmueller M, Morgan H, Scholl I, Rusin V, Daw A, Johnson J, Arndt
  M (2010) Total solar eclipse observations of hot prominence shrouds.
  Astrophys J 719:1362

\bibitem[{Hudson and Schwenn(2000)}]{hudson_00}
Hudson HS, Schwenn R (2000) Hot cores in coronal filament cavities. Adv Space
  Res 25:1859

\bibitem[{Hudson et~al(1999)Hudson, Acton, Harvey, and McKenzie}]{hudson_99}
Hudson HS, Acton LW, Harvey KA, McKenzie DM (1999) A stable filament cavity
  with a hot core. Astrophys J 513:83

\bibitem[{Van~de Hulst(1950)}]{vandehulst}
Van~de Hulst HC (1950) The electron density of the solar corona. Bulletin of
  the Astronomical Institutes of the Netherlands 11:135

\bibitem[{Illing and Hundhausen(1986)}]{illhund}
Illing RM, Hundhausen AJ (1986) Disruption of a coronal streamer by an eruptive
  prominence and coronal mass ejection. J Geophys Res 91:10,951

\bibitem[{{Janse} et~al(2010){Janse}, {Low}, and {Parker}}]{janse_10}
{Janse} AM, {Low} BC, {Parker} EN (2010) {Topological complexity and tangential
  discontinuity in magnetic fields}. Physics of Plasmas 17(9):092,901,
  \doi{10.1063/1.3474943}

\bibitem[{Judge et~al(2006)Judge, Low, and Casini}]{judge_06}
Judge PG, Low BC, Casini R (2006) Spectral lines for polarization measurements
  of the coronal magnetic field. iv. stokes signals in current-carrying fields.
  Astrophys J 651:1229

\bibitem[{Karpen(2014)}]{karpen14_book}
Karpen J (2014) Plasma structure and dynamics. In: Engvold, Vial (eds) Solar
  Prominences, Springer, p~00

\bibitem[{{Klimchuk} et~al(2010){Klimchuk}, {Karpen}, and
  {Antiochos}}]{klimchuk_10}
{Klimchuk} JA, {Karpen} JT, {Antiochos} SK (2010) {Can Thermal Nonequilibrium
  Explain Coronal Loops?} Astrophys J 714:1239--1248,
  \doi{10.1088/0004-637X/714/2/1239}, \eprint{0912.0953}

\bibitem[{{Krall} and {Chen}(2005)}]{krallchen_05}
{Krall} J, {Chen} J (2005) {Density Structure of a Preeruption Coronal Flux
  Rope}. Astrophys J 628:1046--1055, \doi{10.1086/430810}

\bibitem[{Kucera(2014)}]{kucera14_book}
Kucera T (2014) In: Engvold, Vial (eds) Solar Prominences, Springer, p~00

\bibitem[{Kucera et~al(2012)Kucera, Gibson, Schmit, Landi, and
  Tripathi}]{kucera_12}
Kucera TA, Gibson SE, Schmit DJ, Landi E, Tripathi D (2012) Temperature and euv
  intensity in a coronal prominence cavity. Astrophys J 757:73

\bibitem[{Kundu et~al(1978)Kundu, Fuerst, Hirth, and Butz}]{kundu}
Kundu MR, Fuerst E, Hirth W, Butz M (1978) Astron Astrophys 62:431

\bibitem[{{Li} et~al(2012){Li}, {Morgan}, {Leonard}, and {Jeska}}]{li_12}
{Li} X, {Morgan} H, {Leonard} D, {Jeska} L (2012) {A Solar Tornado Observed by
  AIA/SDO: Rotational Flow and Evolution of Magnetic Helicity in a Prominence
  and Cavity}. Astrophys J Lett 752:L22, \doi{10.1088/2041-8205/752/2/L22}

\bibitem[{Liu et~al(2007)Liu, Alexander, and Gilbert}]{liu_07}
Liu R, Alexander D, Gilbert HR (2007) Kink-induced catastrophe in a coronal
  eruption. Astrophys J 661:1260

\bibitem[{{Liu} et~al(2012){Liu}, {Berger}, and
  {Low}}]{LiuW.Berger.Low.flmt-condense.2012ApJ...745L..21L}
{Liu} W, {Berger} TE, {Low} BC (2012) {First SDO/AIA Observation of Solar
  Prominence Formation Following an Eruption: Magnetic Dips and Sustained
  Condensation and Drainage}. Astrophys J Lett 745:L21,
  \doi{10.1088/2041-8205/745/2/L21}, \eprint{1201.0811}

\bibitem[{{Liu} and {Schuck}(2012)}]{liu_12}
{Liu} Y, {Schuck} PW (2012) {Magnetic Energy and Helicity in Two Emerging
  Active Regions in the Sun}. Astrophys J 761:105,
  \doi{10.1088/0004-637X/761/2/105}

\bibitem[{Lopez-Ariste(2014)}]{lopez14_book}
Lopez-Ariste A (2014) In: Engvold, Vial (eds) Solar Prominences, Springer, p~00

\bibitem[{{Low}(1994)}]{low_94}
{Low} BC (1994) {Magnetohydrodynamic processes in the solar corona: Flares,
  coronal mass ejections, and magnetic helicity}. Physics of Plasmas
  1:1684--1690, \doi{10.1063/1.870671}

\bibitem[{Low and Hundhausen(1995)}]{lowhund}
Low BC, Hundhausen JR (1995) Magnetostatic structures of the solar corona. ii.
  the magnetic topology of quiescent prominences. Astrophys J 443:818

\bibitem[{{Low} et~al(1982){Low}, {Munro}, and {Fisher}}]{low_82}
{Low} BC, {Munro} RH, {Fisher} RR (1982) {The initiation of a coronal
  transient}. Astrophys J 254:335--342, \doi{10.1086/159737}

\bibitem[{{Low} et~al(2012{\natexlab{a}}){Low}, {Berger}, {Casini}, and
  {Liu}}]{low_12a}
{Low} BC, {Berger} T, {Casini} R, {Liu} W (2012{\natexlab{a}}) {The
  Hydromagnetic Interior of a Solar Quiescent Prominence. I. Coupling between
  Force Balance and Steady Energy Transport}. Astrophys J 755:34,
  \doi{10.1088/0004-637X/755/1/34}, \eprint{1203.1056}

\bibitem[{{Low} et~al(2012{\natexlab{b}}){Low}, {Liu}, {Berger}, and
  {Casini}}]{low_12b}
{Low} BC, {Liu} W, {Berger} T, {Casini} R (2012{\natexlab{b}}) {The
  Hydromagnetic Interior of a Solar Quiescent Prominence. II. Magnetic
  Discontinuities and Cross-field Mass Transport}. Astrophys J 757:21,
  \doi{10.1088/0004-637X/757/1/21}

\bibitem[{Lugaz(2014)}]{lugaz14_book}
Lugaz N (2014) In: Engvold, Vial (eds) Solar Prominences, Springer, p~00

\bibitem[{{Luna} et~al(2012){Luna}, {Karpen}, and
  {DeVore}}]{Luna.Karpen.promin.thread.HD.2012ApJ...746...30L}
{Luna} M, {Karpen} JT, {DeVore} CR (2012) {Formation and Evolution of a
  Multi-threaded Solar Prominence}. Astrophys J 746:30,
  \doi{10.1088/0004-637X/746/1/30}, \eprint{1201.3559}

\bibitem[{Mackay(2014)}]{mackay14_book}
Mackay D (2014) In: Engvold, Vial (eds) Solar Prominences, Springer, p~00

\bibitem[{Mari\v{c}i\'{c} et~al(2009)Mari\v{c}i\'{c}, Vr\v{s}nak, and
  Rosa}]{maricic_09}
Mari\v{c}i\'{c} D, Vr\v{s}nak B, Rosa D (2009) Relative kinematics of the
  leading edge and the prominence in coronal mass ejections. Solar Phys 260:177

\bibitem[{Marqu\'e(2004)}]{marque_04}
Marqu\'e C (2004) Radiometric observations of quiescent filament cavities.
  Astrophys J 602:1037

\bibitem[{Martin(2014)}]{martin14_book}
Martin S (2014) In: Engvold, Vial (eds) Solar Prominences, Springer, p~00

\bibitem[{{McCabe} and {Mickey}(1981)}]{mccabemickey_81}
{McCabe} MK, {Mickey} DL (1981) {The He I 10,830 A chromosphere and filament
  associated structures}. Solar Phys 73:59--66, \doi{10.1007/BF00153144}

\bibitem[{{McIntosh} et~al(1976){McIntosh}, {Krieger}, {Nolte}, and
  {Vaiana}}]{mcintosh_76}
{McIntosh} PS, {Krieger} AS, {Nolte} JT, {Vaiana} G (1976) {Association of
  X-ray arches with chromospheric neutral lines}. Solar Phys 49:57--77,
  \doi{10.1007/BF00221485}

\bibitem[{{Munro} et~al(1979){Munro}, {Gosling}, {Hildner}, {MacQueen},
  {Poland}, and {Ross}}]{munro_79}
{Munro} RH, {Gosling} JT, {Hildner} E, {MacQueen} RM, {Poland} AI, {Ross} CL
  (1979) {The association of coronal mass ejection transients with other forms
  of solar activity}. Solar Phys 61:201--215, \doi{10.1007/BF00155456}

\bibitem[{{Panasenco} et~al(2014){Panasenco}, {Martin}, and
  {Velli}}]{panasenco_13}
{Panasenco} O, {Martin} SF, {Velli} M (2014) {Apparent Solar Tornado-Like
  Prominences}. Solar Phys 289:603--622, \doi{10.1007/s11207-013-0337-1},
  \eprint{1307.2303}

\bibitem[{Parker(1994)}]{parker}
Parker EN (1994) Oxford University Press, New York

\bibitem[{Rachmeler et~al(2013)Rachmeler, Gibson, Dove, DeVore, and
  Fan}]{rachmeler_13}
Rachmeler LA, Gibson SE, Dove JB, DeVore CR, Fan Y (2013) Polarimetric
  properties of flux ropes and sheared arcades in coronal prominence cavities.
  Solar Phys in press, Arxiv 13047594

\bibitem[{{Reeves} et~al(2012){Reeves}, {Gibson}, {Kucera}, and
  {Hudson}}]{reeves_12}
{Reeves} KK, {Gibson} SE, {Kucera} TA, {Hudson} HS (2012) {Thermal properties
  of coronal cavities observed with the X-ray telescope on Hinode}. Astrophys J
  746:146

\bibitem[{{R{\'e}gnier} et~al(2011){R{\'e}gnier}, {Walsh}, and
  {Alexander}}]{regnier_11}
{R{\'e}gnier} S, {Walsh} RW, {Alexander} CE (2011) {A new look at a polar crown
  cavity as observed by SDO/AIA. Structure and dynamics}. Astron Astrophys
  533:L1, \doi{10.1051/0004-6361/201117381}, \eprint{1107.3451}

\bibitem[{Saito and Hyder(1968)}]{saitohyder}
Saito K, Hyder C (1968) Solar Phys 5:61

\bibitem[{Saito and Tandberg-Hanssen(1973)}]{saitotand_73}
Saito K, Tandberg-Hanssen E (1973) The arch systems, cavities, and prominences
  in the helmet streamer observed at the solar eclipse, november 12, 1966.
  Solar Phys 31:105

\bibitem[{{Schmahl}(1979)}]{schmahl_79}
{Schmahl} EJ (1979) {The prominence-corona interface - A review}. In: {Jensen}
  E, {Maltby} P, {Orrall} FQ (eds) IAU Colloq. 44: Physics of Solar
  Prominences, pp 102--120

\bibitem[{{Schmit} and {Gibson}(2013)}]{schmit_13a}
{Schmit} DJ, {Gibson} S (2013) {Diagnosing the Prominence-Cavity Connection}.
  Astrophys J 770:35, \doi{10.1088/0004-637X/770/1/35}, \eprint{1304.7595}

\bibitem[{Schmit and Gibson(2011)}]{schmit_11}
Schmit DJ, Gibson SE (2011) Forward modeling cavity density: A multi-instrument
  diagnostic. Astrophys J 733:1

\bibitem[{Schmit and Gibson(2014)}]{schmit_14}
Schmit DJ, Gibson SE (2014) The formation of a cavity in a 3d flux rope. IAU
  S300 proceedings In press

\bibitem[{Schmit et~al(2009)Schmit, Gibson, Tomczyk, Reeves, Sterling, Brooks,
  Williams, and Tripathi}]{schmit_09}
Schmit DJ, Gibson SE, Tomczyk S, Reeves KK, Sterling AC, Brooks DH, Williams
  DR, Tripathi D (2009) Large-scale flows in prominence cavities. Astrophys J
  Lett 700:96

\bibitem[{{Schmit} et~al(2013){Schmit}, {Gibson}, {Luna}, {Karpen}, and
  {Innes}}]{schmit_13b}
{Schmit} DJ, {Gibson} S, {Luna} M, {Karpen} J, {Innes} D (2013) {Prominence
  Mass Supply and the Cavity}. Astrophys J 779:156,
  \doi{10.1088/0004-637X/779/2/156}, \eprint{1311.2382}

\bibitem[{{Serio} et~al(1978){Serio}, {Vaiana}, {Godoli}, {Motta},
  {Pirronello}, and {Zappala}}]{serio_78}
{Serio} S, {Vaiana} GS, {Godoli} G, {Motta} S, {Pirronello} V, {Zappala} RA
  (1978) {Configuration and gradual dynamics of prominence-related X-ray
  coronal cavities}. Solar Phys 59:65--86, \doi{10.1007/BF00154932}

\bibitem[{Straka et~al(1975)Straka, Papagiannis, and Kogut}]{straka}
Straka RM, Papagiannis MD, Kogut JA (1975) Solar Phys 45:131

\bibitem[{Tandberg-Hanssen(1974)}]{tand74}
Tandberg-Hanssen E (1974) Solar Prominences. D. Reidel Pub. Co., Dordrecht,
  Holland

\bibitem[{Tandberg-Hanssen(1995)}]{tand95}
Tandberg-Hanssen E (1995) The Nature of Solar Prominences. Kluwer Academic
  Press., Dordrecht, Holland

\bibitem[{Taylor(1974)}]{taylor_74}
Taylor JB (1974) Relaxation of toroidal plasma and generation of reverse
  magnetic fields. Phys Rev Lett 33:19

\bibitem[{{Tian} et~al(2013){Tian}, {Tomczyk}, {McIntosh}, {Bethge}, {de Toma},
  and {Gibson}}]{tian_13}
{Tian} H, {Tomczyk} S, {McIntosh} SW, {Bethge} C, {de Toma} G, {Gibson} S
  (2013) {Observations of Coronal Mass Ejections with the Coronal Multichannel
  Polarimeter}. SP 288:637--650, \doi{10.1007/s11207-013-0317-5},
  \eprint{1303.4647}

\bibitem[{Titov(2007)}]{Titov07}
Titov VS (2007) {Generalized Squashing Factors for Covariant Description of
  Magnetic Connectivity in the Solar Corona}. Astrophys J 660:863

\bibitem[{Titov and Demoulin(1999)}]{titdem}
Titov VS, Demoulin P (1999) Basic topology of twisted magnetic configurations
  in solar flares. Astron Astrophys 351:707

\bibitem[{{T{\"o}r{\"o}k} et~al(2011){T{\"o}r{\"o}k}, {Panasenco}, {Titov},
  {Miki{\'c}}, {Reeves}, {Velli}, {Linker}, and {De Toma}}]{torok_11}
{T{\"o}r{\"o}k} T, {Panasenco} O, {Titov} VS, {Miki{\'c}} Z, {Reeves} KK,
  {Velli} M, {Linker} JA, {De Toma} G (2011) {A Model for Magnetically Coupled
  Sympathetic Eruptions}. Astrophys J Lett 739:L63,
  \doi{10.1088/2041-8205/739/2/L63}, \eprint{1108.2069}

\bibitem[{{Vaiana} et~al(1973){Vaiana}, {Krieger}, and {Timothy}}]{vaiana_73}
{Vaiana} GS, {Krieger} AS, {Timothy} AF (1973) {Identification and Analysis of
  Structures in the Corona from X-Ray Photography}. Solar Phys 32:81--116,
  \doi{10.1007/BF00152731}

\bibitem[{Vasquez et~al(2009)Vasquez, Frazin, and Karmalabadi}]{vasquez09}
Vasquez AM, Frazin RA, Karmalabadi F (2009) 3d temperatures and densities of
  the solar corona via multi-spacecraft euv tomography: Analysis of prominence
  cavities. Solar Phys 256:73

\bibitem[{{Waldmeier}(1970)}]{waldmeier_70}
{Waldmeier} M (1970) {The Structure of the Monochromatic Corona in the
  Surroundings of Prominences}. Solar Phys 15:167--175,
  \doi{10.1007/BF00149483}

\bibitem[{Wang and Stenborg(2010)}]{wang_10}
Wang YM, Stenborg G (2010) Spinning motions in coronal cavities. Astrophys J
  Lett 719:181

\bibitem[{Webb(2014)}]{webb14_book}
Webb D (2014) Eruptive prominences and their impact on the earth; the
  association with coronal mass ejections. In: Engvold, Vial (eds) Solar
  Prominences, Springer, p~00

\bibitem[{{Woltjer}(1958)}]{woltjer_58}
{Woltjer} L (1958) {A Theorem on Force-Free Magnetic Fields}. Proceedings of
  the National Academy of Science 44:489--491, \doi{10.1073/pnas.44.6.489}

\bibitem[{Zhang and Low(2005)}]{zhanglow_05}
Zhang M, Low BC (2005) The hydromagnetic nature of solar coronal mass
  ejections. Ann Rev of Astron and Astrophys 43:103

\end{thebibliography}
\end{document}